\documentclass[letterpaper,11pt,onecolumn,oneside,notitlepage,openbib,final]{article}

\usepackage{mathpazo}	
\usepackage{relsize}
\renewcommand{\ast}{{\mathlarger *}} 
\usepackage{float}
\usepackage{lscape}	
\usepackage[bottom]{footmisc}			
\usepackage[margin=1in]{geometry}
\usepackage{setspace} 
\usepackage{amsmath,amsfonts,amssymb,amsthm}
\usepackage{bbm} 
\usepackage{mathrsfs}		
\allowdisplaybreaks
\usepackage[pdflatex]{graphicx}
\usepackage{color}
\usepackage[labelformat=parens]{subfig}		
\usepackage[notref,notcite,color]{showkeys}
\definecolor{labelkey}{gray}{.30}
\renewcommand*\showkeyslabelformat[1]{%
  \parbox[t]{\marginparwidth}{\raggedright\scriptsize\path{#1}}}
\usepackage[pdftex,bookmarks=true,bookmarksnumbered=true,colorlinks=true,linkcolor=black,%
	citecolor=black,filecolor=black,urlcolor=black]{hyperref}
\usepackage{url}
\usepackage{natbib}
\usepackage{enumerate}
\usepackage{alltt}	
\usepackage{cleveref} 

\def\pdif#1#2{\frac{\partial #1}{\partial #2}}

\def\2'{^{\prime\prime}}

\def\bone{\mathbbm 1}

\def\E{\mathbb E}
\def\e{\mathbf e}

\def\F{\mathbf F}
\def\g{\mathbf g}
\def\m{\mathbf m}
\def\M{\mathbf M}
\def\N{\mathbb N}
\def\P{\mathbb P}

\def\R{\mathbb R}

\def\uniqBR{\text{uniqBR}}
\def\BR{\text{BR}}
\def\s{\mathfrak s}
\def\V{\mathbf V}
\def\v{\mathbf v}
\def\X{\mathbf X}
\def\x{\mathbf x}
\def\y{\mathbf y}
\def\Y{\mathbf Y}
\def\z{\mathbf z}
\def\0{\mathbf 0}
\def\1{\mathbf 1}
\def\O{\mathbf O}

\def\cA{\mathcal A}
\def\cB{\mathcal B}
\def\cC{\mathcal C}
\def\cF{\mathcal F}

\def\cM{\mathcal M}
\def\cX{\mathcal X}

\def\cS{\mathcal S}

\def\bpi{{\boldsymbol{\pi}}}
\def\brho{{\boldsymbol{\rho}}}
\def\btheta{{\boldsymbol{\theta}}}
\def\bmu{{\boldsymbol{\mu}}}

\renewcommand{\mid}{~:~}

\DeclareMathOperator*{\argmax}{argmax}

\DeclareMathOperator{\cl}{cl}

\theoremstyle{plain}
\newtheorem{thm}{Theorem}\crefname{thm}{theorem}{theorems}
\newtheorem{lem}{Lemma}\crefname{lem}{lemma}{lemmas}

\newtheorem{cor}{Corollary}\crefname{cor}{corollary}{corollaries}

\theoremstyle{definition}
\newtheorem{dfn}{Definition}\crefname{dfn}{definition}{definitions}

\newtheorem{assmp}{Assumption}\crefname{assmp}{assumption}{assumptions}

\theoremstyle{remark}
\newtheorem{remk}{Remark}

\newtheorem{exmpl}{Example}

\newcounter{exmp}\crefname{exmpl}{example}{examples}
\newenvironment{exmp}[1][]{\stepcounter{exmp}\begin{exmpl}[#1]}{\hfill \rule{0.33em}{0.8em}\end{exmpl}}

\newenvironment{exmppr}[2][]
  {\renewcommand{\theexmpl}{\ref{#2}$'$}%
   \addtocounter{exmpl}{-1}%
   \begin{exmpl}[#1]}
  {\hfill \rule{0.33em}{0.8em}\end{exmpl}}

\newenvironment{rmk}[1][]{\begin{remk}[#1]}{\hfill \rule{0.33em}{0.8em}\end{remk}}

\makeatletter 
\long\def\@makecaption#1#2{
 \vskip 2pt 
 \setbox\@tempboxa\hbox {\small {\it #1:} #2} 
 \ifdim \wd\@tempboxa >\hsize {\small {\it #1}: #2\par} \else \hbox
to\hsize{\hfil\box\@tempboxa\hfil} 
 \fi}
\makeatother

\title{Evolutionary dynamics in heterogeneous populations: \\ a general framework for an arbitrary type distribution}
\author{Dai {\sc Zusai}\thanks{Department of Economics, Temple University, 1301 Cecil B. Moore Ave., RA 873 (004-04), Philadelphia, PA 19122, U.S.A. Tel:+1-215-204-8880. E-mail: \texttt{zusai@temple.edu}. The supplementary note is available from the author's website: \url{https://sites.temple.edu/zusai/research/disaggevol/}.} }
\date{April 7, 2019}

\begin{document}
\setlength{\abovedisplayskip}{0.9ex}
\setlength{\belowdisplayskip}{0.9ex}
\maketitle

\begin{abstract}
We present a general framework of evolutionary dynamics under persistent heterogeneity in payoff functions and revision protocols,  allowing continuously many types in a game with finitely many strategies. The dynamic is rigorously formulated as a differential equation of a joint probability measure of types and strategies. To establish a foundation of this framework, we clarify regularity assumptions on the revision protocol, the game and the type distribution to guarantee the existence of a unique solution trajectory as well as those to guarantee the existence of equilibrium in a heterogeneous population game. We further verify equilibrium stationarity in general and stability in potential games under admissible dynamics. 


\end{abstract}
\newpage
\section{Introduction}\label{sec:intro}
Evolutionary dynamics formulate off-equilibrium adjustment processes of agents' choices in games, allowing various decision rules  (revision protocols) such as exact optimization, better reply based on pairwise comparison of payoffs, imitation, etc. Despite a wide range of applications to social and economic problems and also a potential role to challenge a conventional equilibrium-based approach, evolutionary dynamics have not fully captured one common staple of economic/agent-based models, i.e., heterogeneity of agents. It is a common practice in applied or empirical studies to assume continuous types of agents---especially, in many of applied economic models (e.g. auctions, aggregate demand\footnote{\label{ftnt:DynPricing}Dynamic demand of myopic consumers is considered in the literature on dynamic monopoly pricing: \cite{Rohlfs74Bell,DhebarOren85MS,DhebarOren86OR} are seminar papers. They assume a continuous type distribution to define a continuous dynamic of the aggregate demand, though they implicitly assume aggregability. Employing the aggregability result in \cite{ElySandholmBayes1}, \cite{Zusai_Platform} justifies the aggregate demand dynamic as an aggregate obtaiend from the standard best response dynamic.}), in econometric estimation of discrete choice models (e.g. logit regression) and in theoretical investigations of game experiments (quantal response equilibria). To embed heterogeneity to evolutionary dynamics, we typically assume that there are only finitely many types so they can be formulated as distinct populations (or genes); it requires some technical twists for discrete approximation of a continuous type space and also leaves non-negligible impacts of each individual type on others.\footnote{For example, \cite{Lahkar17_LogitDyn_AggPotGame_MultiTypes} considers logit dynamics in a potential game on a continuous strategy space with finitely many payoff types.}

There are a few studies that deal with a continuous range of payoff heterogeneity in evolutionary dynamics. But, these studies rely on \textit{aggregability} of the dynamic---the change in the aggregate strategy distribution is wholly determined from the current state of the aggregate distribution alone, independently of the underlying correlation between strategy choices and payoff types.\footnote{ \cite{Hummel_McAfee_18IER_EvolCons_MonopExit} apply (a generalized version of) replicator dynamics to formulate the demand dynamic in the monopoly pricing problem, as argued in footnote \ref{ftnt:DynPricing}. While the replicator dynamic is not aggregable as argued in \cite{ZusaiDistStbl}, they obtain an explicit solution for the differential equation that represent the demand dynamic, thanks to their specification of functions (especially in their Lemma 1). Since the demand dynamic is only a part of the monopolist's dynamic optimization, equilibrium stationarity or stability is not discussed in their paper. (Actually, terms like `equilibrium' or 'stability' appears only in the bibliography in the paper.)} Such aggregability may be assumed as in \cite{Blonski99GEB_AnonymousGame_binary} or may be derived from some specific form of the agents' strategy revision processes as in \cite{ElySandholmBayes1}. This is a demanding restriction for games and dynamics; heterogeneous choices of agents cannot have an impact on payoffs or on dynamics through something beyond their average, for example through the variance or distribution of strategies over different types. It is virtually the same as having just \textit{one} ``representative/average'' type of agents. This cannot capture impacts of the correlation between persistent heterogeneity (or ``fixed effects'' in discrete choice regression) and changes in choices.

In this paper, we provide a general framework to extend evolutionary dynamics to heterogeneous population games without requiring aggregability or restricting to a finite type space. We allow agents not only to have different payoff functions but also to follow different decision rules. To allow continuously many types in our framework, we need to tackle on technicality on dimensionality. The state of an evolutionary dynamic is the strategy distribution over different types; the dimension of the dynamic is just as large as the number of types. Without averaging off heterogeneity or assuming a finite type space, we need to deal with a dynamic system on infinite dimension. Therefore, we start from carefully defining evolutionary dynamics with a measure theoretic formulation of the state space, borrowed from the literature on evolution in games with  continuously many \textit{strategies}, especially   \cite{OechsslerRiedelET01_InfStrEvolDyn,OechsslerRiedelJET02_EvolStbl_Cont} and  \cite{Cheung13_PairwiseCompDyn_ContStr}.\footnote{To name a few more, see also \cite{HofOechsslerRiedel09GEB_BvNDyn_ContStr}, \cite{FriedmanOstrov13JET_EvolDyn_ContsActions}, \cite{LahkarSeymour13GEB_Reinforcmt_PopGames},  \cite{LahkarRiedel15GEB_LogitDyn_ContStr} and \cite{Cheung16GEB_ImitatDyn_ContStr}.} 

Even the unique existence of a solution trajectory cannot be simply granted for infinite dimensional dynamics. We clarify the regularity conditions on games and individual decision rules to assure it. If individual agents respond continuously to changes in payoffs (L-continuous revision protocols in \Cref{dfn:ContRev}), the dynamic has a unique solution trajectory from an arbitrary initial state in the heterogeneous setting, requiring only the uniform boundedness of switching rates over all types (\Cref{ass:bddR}). If an agent takes only the exact optimal strategy (exact optimization protocols in \Cref{dfn:ExactOpt}) just as in the best response dynamic, the individual revision protocol exhibits discontinuity when the transition of the strategy distribution triggers a switch of the agent's optimal strategy through changes in payoffs. To mitigate discontinuity at the individual level, we impose a kind of Lipschitz continuity on the distribution of the types whose best response strategies change with such a transition (\Cref{ass:cont_type}).

We then confirm that standard properties of evolutionary dynamics can be extended from the homogeneous setting to the heterogeneous setting. First, if the individual decision rule assures stationarity of Nash equilibrium in the homogeneous setting, it also assures equilibrium stability in heterogeneous population games (\Cref{thm:NStat}). We also obtain the condition for the existence of an equilibrium (\Cref{thm:ExistEqm}). Combining them, we can guarantee existence of a stationary state in heterogeneous evolutionary dynamics. While stability of equilibrium is not granted generally even in a homogeneous population game, it is known that potential games maintain equilibrium stability over a wide range of homogeneous evolutionary dynamics. With a rigorous formulation of heterogeneous potential games (\Cref{dfn:Pot_Hetero}), we verify that equilibrium stability is extended to the heterogeneous setting (\Cref{thm:NStbl_pot}). In particular, a local maximum of the potential function is locally stable under \textit{any} admissible dynamics, and also vice versa if the maximum is isolated (\Cref{cor:NStbl_pot_SpecifDyn}). 

In the next section, we define a heterogeneous population game and then build a heterogeneous evolutionary dynamic from an individual agent's revision protocol. Next we present our main results. In \Cref{sec:Exist}, we study the regularity conditions to guarantee the existence of a unique solution path. In \Cref{sec:EqmComp_Ext}, we extend equilibrium stationarity in general and equilibrium stability of potential games to the heterogeneous setting. Until this section, we consider heterogeneity only in payoff functions and focus on non-observational evolutionary dynamics, in which an agent's switching rate depends only on the payoff vector for the agent but not on other agents' strategies. In \Cref{sec:ext}, we consider heterogeneity in revision protocols and observational dynamics such as imitative dynamics and excess payoff comparison dynamics; we confirm that the theorems in this paper are robust to these extensions. We conclude the paper in \Cref{sec:concl} with a summary of the positive results in this paper and discussion on their implications and limitations. 
Appendices provide the proofs and a few technical details on the measure-theoretic construction of heterogeneous dynamics. Parts of proofs that essentially involve only heavy calculation are found in the Supplementary Note.

\section{The base model}\label{sec:Model}
\subsection{Heterogeneous population games }\label{sec:game}
We first set up the game played in a heterogeneous population; here we quickly introduce essential components for our analysis, while we provide a complete measure-theoretic formulation in Appendix  \ref{sec:comp_Bayes}. 

Consider a continuous population of agents, each of whom chooses a strategy from the same strategy set $\cS=\{1,\cdots,S\}.$ 
Each agent is assigned to type $\btheta\in\Theta$. Assume that type space $\Theta$ is a complete separable space with metric $d_\Theta:\Theta^2\to\R_+$.\footnote{In Examples \ref{exmp:ASAG_Dfn}-- \ref{exmp:RndMat_TransitSig}, $\Theta$ is (possibly, a subset of) a finite-dimensional real space such as $\R$ or $\R^S$. Then, we can take an Euclidean norm, for example, to define metric $d_\Theta$.} Types may represent heterogeneity in assessments of payoffs (possibly due to private information) as we focus in this base model, or heterogeneity in revision protocols as we discuss in \Cref{sec:ext}, or both. If there are only finitely many types, these ``types'' could be formulated as different populations (or species in  a biological context) in a conventional approach; however, we may have continuously many types in our model. Let $\cB_\Theta$ the set of Borel sets over $\Theta$, and $\P_\Theta$ be the distribution of types: for any Borel set $B_\Theta\in\cB_\Theta$ of types, $\P_\Theta(B_\Theta)$ is the mass of agents whose types belong to $B_\Theta$.

The population state is described by the \textbf{(joint) strategy distribution} $\X=(X_s)_{s\in\cS}$, a joint distribution of strategies and types such that the marginal distribution of types coincides with $\P_\Theta$. For each strategy $s\in\cS$ and each Borel set $B_\Theta\in\cB_\Theta$ of types, $X_s(B_\Theta)$ is a mass of strategy-$s$ players whose types belong to $B_\Theta$. For each $B_\Theta$, the strategy distribution $\X=(X_s)_{s\in\cS}$ must satisfy $\sum_{s\in\cS} X_s(B_\Theta)=\P_\Theta(B_\Theta).$ Denote by $\cX$ the space of joint strategy distributions.

Since $\X$ satisfies $X_s(B_\Theta)\le \P_\Theta(B_\Theta)$ for each $s\in\cS$, each $X_s$ is absolutely continuous with respect to $\P_\Theta$; see \eqref{eq:dfn_sbscont} in \Cref{sec:comp_Bayes}. We denote this relationship of the absolute continuity by $\P_\Theta\gg \X$. By Radon-Nikodym theorem, the absolute continuity guarantees the existence of a density function $x_s:\Theta\to\R_+$ of $X_s$ such that $X_s(B_\Theta)=\int_{B_\Theta} x_s d\P_\Theta$.  Then, \textbf{(type-)conditional strategy distribution} $\x=(x_s)_{s\in\cS}$ is defined by collecting the density functions $x_s$ over all $s\in\cS$;\footnote{In an incomplete information game with finite players, $\X$ is essentially a distributional strategy and $\x$ is a behavioral strategy in \cite{MilgromWeber85MathOR_DistrStr}. \cite{ElySandholmBayes1} call $\x$ a Bayesian strategy.} we abbreviate the relationship between $\X$ and its density $\x$ as $\X=\int \x d\P_\Theta$. Notice $\x(\btheta)\in\Delta^\cS:=\{\z\in\R^S_+ : \sum_{s\in\cS} z_s=1\}$ for each type $\btheta\in\Theta$.\footnote{We denote $\R_+=[0,+\infty)$ and $\R_{++}=(0,+\infty)$. Consider a $|\mathscr U|$-dimensional real space, each of whose coordinate is labeled with one element of $\mathscr U=\{1,\ldots,|\mathscr U|\}$. For set $S\subset \mathscr U$, we define an $|S|$-dimensional simplex $\Delta^{|\mathscr U|}(S)$ as $\Delta^{\mathscr U}(S):=\left\{ \x\in\R^{|\mathscr U|}_+ \mid \sum_{k\in S}x_k=1 \text{ and }x_l=0 \text{ for any }l\in\mathscr U\setminus S \right\}$. When $S$ is the whole space $\mathscr U$ itself, we omit $|\mathscr U|$ and denote it by $\Delta^{\mathscr U}$.} $x_s(\theta)\in[0,1]$ can be interpreted as the population share of strategy-$s$ players in the subpopulation of type-$\btheta$ agents. Denote by $\cF_\cX$ the set of type-conditional strategy distributions.\footnote{Two behavioral strategies $\x,\x'\in\cF_\cX$ are considered as identical, i.e., $\x=\x'$ if $\x(\btheta)=\x'(\btheta)$ for $\P_\Theta$-almost all $\btheta\in\Theta$. They indeed yield the same joint strategy distribution.} Type-conditional strategy distribution $\x$ is ($\P_\Theta$-almost) uniquely determined from joint strategy distribution $\X$ by Radon-Nikodym theorem, and vice versa. So, $\cX$ is equivalent to $\cF_\cX$.

Let $F_s[\X](\btheta)$ be a type $\btheta$-agent's payoff from strategy $s$ when the strategy distribution is $\X$. Thus, $\F[\X](\btheta)=(F_s[\X](\btheta))_{s\in\cS}\in\R^S$ is the payoff vector  for type $\btheta$ given strategy distribution $\X$. Given $\X$, $\F[\X]:\Theta\to\R^S$ specifies the payoff vector $\F[\X](\btheta)$ for each type $\btheta\in\Theta$; thus we call $\F[\X]$ the payoff vector profile. We assume that $\F[\X]$ belongs to $\cC_\Theta$, the set of continuous functions from $\Theta$ to $\R^S$. Payoff function $\F:\cX\to\cC_\Theta$ maps a strategy distribution $\X\in\cX$ to a payoff vector profile $\F[\X]\in\cC_\Theta$. A \textbf{heterogeneous population game} is defined by $(\cS,(\Theta,\cB_\Theta,\P_\Theta),\F)$, which we represent by $\F$. 



Let $\cS_\BR(\bpi^0)\subset\cS$ be the set of optimal strategies given payoff vector $\bpi^0=(\pi^0_s)_{s\in\cS}\in\R^S$: i.e., $ \cS_\BR(\bpi^0):=\argmax_{s\in\cS} \pi^0_s.$ Denote by $\Delta( \cS_\BR(\bpi^0))$ the set of strategy distributions that assign positive probabilities only to the optimal strategies given $\bpi^0$: i.e., $ \Delta(\cS_\BR(\bpi^0))=\{\y\in\Delta^S : y_s>0 \ \Rightarrow\ a\in \cS_\BR(\bpi^0)\}.$

In heterogeneous population game $\F$, $\cS_\BR^\F[\X](\btheta):=\cS_\BR(\F[\X](\btheta))$ collects the optimal strategies given payoff vector $\F[\X](\btheta)$ for type $\btheta$; namely, it is the set of type-$\btheta$'s best response strategies to $\X$ in game $\F$. Let $\Theta^\F_{s\in\BR}[\X]$ be the set of types for which strategy $s$ is a best response to $\X$, and $\Theta^\F_{s=\uniqBR}[\X]$ the set of types for which strategy $s$ is the \textit{unique }best response to $\X$: i.e., 
$$ \Theta^\F_{s\in\BR}[\X]:=\{\btheta\in\Theta : s\in \cS^\F_\BR [\X](\btheta)\} \quad \supset \quad
\Theta^\F_{s=\uniqBR}[\X]:=\left\{\btheta\in\Theta ~:~ \{s\}=\cS^\F_\BR[\X](\btheta)\right\}. $$

In a Nash equilibrium, (almost) every agent correctly predicts  strategy distribution $\X$ and takes the best response to it. Correspondingly, strategy distribution $\X\in\cX$ is an \textbf{equilibrium strategy distribution} in game $\F$, if
\begin{equation}
\P_\Theta(\Theta^\F_{s=\uniqBR}[\X] \cap B_\Theta)\le X_s(B_\Theta)\le \P_\Theta(\Theta^\F_{s\in\BR}[\X] \cap  B_\Theta) \qquad\text{for all $s\in\cS$ and $B_\Theta\in\cB_\Theta$. }\label{EqmComp}
\end{equation}
Among  types in $B_\Theta$, all those who have $s$ as the \textit{unique }best response \textit{must} choose this strategy $s$ in equilibrium; thus $X_s(B_\Theta)$ must be at least $\P_\Theta(\Theta^\F_{s=\uniqBR}[\X] \cap B_\Theta)$. On the other hand, those who have $s$ as \textit{one }of the best responses \textit{may or may not} add to strategy-$s$ players and thus $X_s(B_\Theta)$ is at most $\P_\Theta(\Theta^\F_{s\in\BR}[\X] \cap B_\Theta)$. In terms of conditional strategy distribution $\x$ such that $\X=\int\x d\P_\Theta$, \eqref{EqmComp} is equivalent to 
\begin{equation}
\x(\btheta)\in \Delta (\cS_\BR^\F[\X](\btheta)) 
\qquad\text{ for $\P_\Theta$-almost all $\btheta\in\Theta$}, \label{BayesEqm}
\end{equation}
or equivalently, 
\begin{equation}
x_s(\btheta)=
	\begin{cases} 
	1 & \text{ if } \btheta\in\Theta^\F_{s=\uniqBR}[\X]\\
	0 &\text{ if } \btheta\notin \Theta^\F_{s\in\BR}[\X]
	\end{cases}  
	\qquad\text{ for all $s\in\cS$ and $\P_\Theta$-almost all $\btheta\in\Theta$}.\tag{\ref{BayesEqm}'}\label{BayesEqm_Sorted}
\end{equation}
That is, if $s$ is the unique best response for type $\btheta$, (almost) all the agents of this type should take it; if $s$ is not a best response, (almost) none of these agents should take it. We leave indeterminacy of $x_s(\btheta)$ in an equilibrium when there are multiple best response strategies for $\btheta$ and $s$ is just one of them.

\subsubsection*{Examples of heterogeneous population games}

\begin{exmp}[Adding payoff heterogeneity to a homogenous population game\label{exmp:ASAG_Dfn}]
Denote by $\bar x_s:=X_s(\Theta)=\E_\Theta x_s\in [0,1]$ the mass of agents who take strategy $s\in\cS$ in the entire population over all types in $\Theta$.\footnote{Here $\E_\Theta$ is the expectation operator on the probability space $(\Theta,\cB_\Theta,\P_\Theta)$: i.e., $\E_\Theta \tilde f:=\int_\Theta \tilde f(\btheta)\P_\Theta(d\btheta)$ for a $\cB_\Theta$-measurable function $\tilde f:\Theta\to\R$.} We call $\bar\x:=(\bar x_s)_{s\in\cS}\in\Delta^\cS$ the \textbf{aggregate strategy distribution}. If each type's payoff function $\F(\btheta):\cX\to\R^S$ depends only on aggregate strategy, that is, $\F$ satisfies $\F[\X](\btheta)=\F[\X'](\btheta)$ for any type $\btheta\in\Theta$ under any pair of two strategy distributions $\X,\X'\in\cX$ that yields the same aggregate strategy distribution $\X(\Theta)=\X(\Theta')$, then we call the game an \textbf{aggregate game}.\footnote{Notice the difference from an \textit{aggregative }game \citep{Corchon94_CompStat_AggGames_StrConcavity,Jensen18Hdbk_AggregativeGames}. The payoff depends only on the population-weighted sum of strategies $\sum_{s\in\cS} s \bar x_s$ in a linearly aggregative game, or a scalar-valued summary $g(\bar\x)\in\R$ in a generalized aggregative game; to make sense, strategies must be some quantities. Cournot competition where strategy $s$ is the quantity of production is a canonical example of an aggregative game. An aggregative game is a special case of aggregate games and wider than it, since it does not require the aggregate strategy distribution $\bar\x\in\R^S$ to reduce to a scalar.}

Especially, in the context of discrete choice models such as in \cite{AndersonDePalmaThisse92_DiscChoice}, it is common to introduce payoff heterogeneity in an additively separable manner. That is, the payoff function is additively separated to the common part and the idiosyncratic part: with type space $\Theta\subset\R^S$, type $\btheta=(\theta_s)_{s\in\cS}\in\R^S$ is defined as the idiosyncratic payoff vector for this type, which varies among agents but does not change over time regardless of the state of the population. Given aggregate state $\bar\x$, $\F^0(\bar\x)=(F^0_s(\bar\x))_{s\in\cS}\in\R^S$ is the common payoff vector, shared by all the agents in the entire population. Thus, at each strategy distribution $\X\in\cX$, the payoff vector for a type-$\btheta$ agent is
\begin{equation}
\F[\X](\btheta)=\F^0(\X(\Theta))+\btheta. \label{eq:F_dfn}
\end{equation}
We call an aggregate game with such additively separable idiosyncratic payoffs an \textbf{additively separable aggregate game (ASAG)}. We can regard an ASAG as an extension of a homogeneous population game $\F^0$ to a heterogeneous setting.
\end{exmp}

\begin{exmp}[Random matching in an incomplete information game\label{exmp:RndMat_IncompInfo}]
An agent is randomly matched with another agent from the same population: let the type space be $\Theta\subset\R$. Given the agent's own type $\theta$ and strategy $s$ and the opponent's type $\theta'$ and strategy $s'$, the agent receives the payoff $u_{ss'}(\theta,\theta')$; let $U(\theta,\theta')=(u_{ss'}(\theta,\theta'))_{(s,s')\in\cS^2}$ the payoff matrix for type $\theta$ matched with $\theta'$. Suppose that an agent cannot observe the opponent's type and thus has to take a strategy independently of the opponent's type. 

We can regard random matching in this two-player incomplete information game as a heterogeneous population game, where the payoff function $\F$ is defined as the expected payoff in the random matching: for a type-$\theta$ agent, the payoff from strategy $s\in\cS$ is\footnote{Here $\e_s\in\Delta^\cS$ is a (column) vector whose coordinates are all 0, except the $s$-th one being 1.}
$$ F_s[\X](\theta)=\int_{\theta'\in\Theta} \sum_{s'\in\cS} u_{ss'}(\theta,\theta') X_{s'}(d\theta)= \int_{\theta'\in\Theta} \e_s\cdot U(\theta,\theta') \X(d\theta').$$
Note that $\F[\X](\theta)=\int_\Theta U(\theta,\theta') \X(d\theta')\in\R^S$.
\end{exmp}

\begin{exmp}[Random matching in an incomplete information game with transitory signals\label{exmp:RndMat_TransitSig}]
In the previous example, an agent's type $\theta$ directly affects the payoff $U$ in the game played after random matching while the agent has no information about the opponent's type. Instead, here we assume that, at each matching, agents receive noisy signals of their types. That is, when a type-$\theta$ agent is matched with a type-$\theta'$ agent, the (former) agent and the opponent receive signals $\hat\theta$ and $\hat\theta'$ respectively with probability $\P_{\hat\Theta}(\hat\theta,\hat\theta'|\theta,\theta')$. Assume that these signals $\hat\theta,\hat\theta'$ are drawn from a finite set $\hat\Theta$ and an agent cannot observe the other agent's signal. To distinguish the strategy set in this post-match two-player game from the one in the random matching population game, denote the former by $\hat\cA$ and assume it is a finite set: an agent chooses an action from $\hat\cA$ after receiving a signal. When the type-$\theta$ agent receives signal $\hat\theta\in \hat\Theta$ and chooses action $\hat a\in\hat\cA$ and its opponent receives signal $\hat\theta' \in \hat\Theta$ and chooses action $\hat a'\in \hat\cA$, the former agent receives payoff $u_{\hat a \hat a'}(\theta,\theta';\hat\theta,\hat\theta')$ and the latter receives $u_{\hat a' \hat a }(\theta',\theta;\hat\theta',\hat\theta)$.

Again, we can regard the random matching as a heterogeneous population game. The ``strategy'' in the random-matching population game should be a complete contingent plan of actions, i.e., a mapping  $s:\hat\Theta\to\hat\cA$ from each possible signal $\hat\theta\in\hat\Theta$ to an action $\hat a\in\hat\cA$ in the post-match two-player game. The ``strategy'' set $\cS$ in the population game is defined as a set of mappings from $\hat\Theta$ to $\hat\cA$; it is a finite set since $\hat\Theta$ and $\hat\cA$ are finite sets. A type-$\theta$ agent's payoff from $s:\hat\Theta\to\hat\cA$ in this population game is defined as an expected payoff such that
$$ F_s[\X](\theta)=\int_{\theta'\in\Theta} \sum_{s'\in\cS}  \sum_{(\hat\theta,\hat\theta')\in \hat\Theta^2} u_{s(\hat\theta) s'(\hat\theta')}(\theta,\theta';\hat\theta,\hat\theta') \P_{\hat\Theta}(\hat\theta, \hat\theta' | \theta,\theta') X_{s'}(d\theta').$$
\end{exmp}

\begin{exmp}[Structured population game]\label{exmp:StrPop}
We could interpret a type just as a ``population" in a conventional model in evolutionary game theory, while we allow continuously many populations. Then, a type represents an affiliation to a certain subgroup of agents in the society; so $\Theta$ is a set of subgroups. Say, a base game is a two-population game $\F^0:\Delta^\cS\times\Delta^\cS \to\R^S$; an agent chooses a strategy, say $s$, from $\cS$ and then receives payoff $F^0_s(\x,\x')$ given the (conditional) strategy distribution in own population $\x\in\Delta^\cS$ and that in the opponent's population $\x'\in\Delta^\cS$. When the society is divided into many subgroups, their connections may not be uniform. Say, an agent in subgroup $\theta$ assigns weight $g(\theta,\theta')$ to the game with subgroup $\theta'$.\footnote{We can allow $g(\theta,\theta')$ to be negative, which implies that an agent has a reversed preference in interactions with subgroup $\theta'$. For example, if a base game is a coordination game, an agent may want to coordinate to the same action with a `friend'; but, with agents in an `enemy' subgroup, the agent wants to take a different action. See Example 1 in \cite{WuZusai19_ConnectedEvol}.} Assuming that an agent must apply the same strategy to any opponent subgroups, the total payoff for an agent in subgroup $\theta$ from strategy $s$ given the joint strategy distribution $\X$ is 
$$ F_s[\X](\theta):= \int_\Theta F^0_s(\x(\theta),\x(\theta'))g(\theta,\theta')\P_\Theta(d\theta').$$
This defines a population game $\F$. \cite{WuZusai19_ConnectedEvol} call such a game a \textit{structured population game}\footnote{They restrict attention to a finitely many subgroups of agents who play a linear game (with no influence of the own population) such as $\F^0(\x,\x')=U^0 \x'$ with an $S\times S$ matrix $U^0$, while they consider both the medium run dynamic where an agent's affiliation is fixed exogenously (as in our model) and the long run dynamic where an agent can change both strategy and affiliation (not covered in this paper).}
\end{exmp}

\subsection{Evolutionary dynamics}\label{sec:dyn}
In an evolutionary dynamic, an agent occasionally changes the strategy over a continuous time horizon $\R_+$, following a Poisson process. The timing of a switch and the choice of which strategy to switch to are determined by \textbf{revision protocol} $\brho=(\rho_{s s'})_{s,s'\in\cS}:\R^S\to \R^{S\times S}_+$, a collection of switching rate functions $\rho_{s s'}:\R^S\to\R_+$ over all the pairs $(s,s')\in \cS\times\cS$ of two strategies. An economic agent should base the switching decision on the payoff vector that the agent is facing. Let $\bpi^0\in\R^S$ be the payoff vector for the agent. The switching rate $\rho_{s s'}(\bpi^0)\in\R_+$ is a Poisson arrival rate at which this agent switches to strategy $s'\in\cS$ conditional on that the agent has been taking strategy $s\in\cS$ so far and currently faces payoff vector $\bpi^0$. The analysis in this paper is applicable to \textit{observational dynamics}, in which the switching rates also depend on the strategy distribution. In addition,  all our theorems hold even when different types of agents follow different revision protocols. We confirm applicability to these extensions in \Cref{sec:ext}, while we focus on heterogeneity only in payoff functions and thus assume that all the types of agents share the same revision protocol $\brho$ until that section. 


In the heterogeneous setting, different types of agents may face different payoff vectors. Let $\bpi:\Theta\to\R^S$ be a payoff vector profile that specifies payoff vector $\bpi(\btheta)$ of each type $\btheta$. From revision protocol $\brho:\R^S\to \R^{S\times S}_+$, we construct the mean dynamic of conditional strategy distribution $\x$ over $\cF_\cX$ with function $\v=(v_s)_{s\in\cS}:\R^S\times\Delta^S\to\R^S$ as 
\begin{equation}
\dot x_s(\btheta) =v_s(\bpi(\btheta),\x(\btheta)):= \sum_{s'\in\cS} x_{s'}(\btheta) \rho_{s' s}(\bpi(\btheta))  - x_s(\btheta)\sum_{s'\in\cS} \rho_{s s'}(\bpi(\btheta)) \label{eq:Dyn_x}
\end{equation}
for each type $\btheta\in\Theta$ and each strategy $s\in\cS$, i.e., $\dot\x(\btheta)=\v(\bpi(\btheta),\x(\btheta))$. In an infinitesimal length of time $dt\in\R$, $\sum_{s'\in\cS} x_{s'}(\btheta) \rho_{s' s}(\bpi(\btheta))dt$ is approximately the mass of type-$\btheta$ agents who switch to strategy $s$ from other strategies $s'\in\cS$, namely, the gross inflow to $x_s(\btheta)$; similarly, $x_s(\btheta)\sum_{s'\in\cS} \rho_{s s'}(\bpi(\btheta))dt$ is the gross outflow from $x_s(\btheta)$. Thus, $v_s(\bpi(\btheta),\x(\btheta))dt$ is the net flow to  $x_s(\btheta)$ in this period of time $dt$.

Coupled with a heterogeneous population game $\F$, the mean dynamic \eqref{eq:Dyn_x} of strategy distribution $\x$ defines the dynamic $\v^\F$ of conditional strategy distribution in $\cF_\cX$ by 
$$ \dot\x(\btheta)=\v^\F[\x](\btheta) := \v(\F[\X](\btheta),\x(\btheta))\in\R^S \quad \text{ 
for each type $\btheta\in\Theta$}, \text{ where }\X=\int \x d\P_\Theta.$$
By collecting $\v^\F[\x]$ over types, we can further define the \textbf{heterogeneous dynamic} of joint strategy distribution in $\cX$ as 
$$\dot\X(B_\Theta)=\V^\F[\X](B_\Theta) := \int_{B_\Theta} \v^\F[\x](\btheta) \P_\Theta(d\btheta) \qquad\text{ for each }B_\Theta\in\cB_\Theta.$$
Note that an agent's type $\btheta$ is \textit{persistently fixed} over time: each agent draws its type $\btheta$ from $\Theta$ at time 0 and keeps it forever.

\subsubsection*{Examples of evolutionary dynamics}
To make a concrete image of revision protocols, here we review major evolutionary dynamics.\footnote{\label{ftnt:SkipExmplDyn}Readers who are familiar with major evolutionary dynamics may just scan this subsection quickly and jump to \Cref{dfn:ContRev,dfn:ExactOpt}.} In particular, we separate the dynamics based on optimization from others because they need different regularity conditions to guarantee the existence of a unique solution trajectory. 

\paragraph*{L-continuous revision protocols.} Under an L-continuous revision protocol $\brho$, the switching rate function $\rho_{s s'}$ is a Lipschitz continuous function of the payoff vector.

\begin{dfn}[L-continuous revision protocols]\label{dfn:ContRev}
In an \textbf{L-continuous revision protocol} $\brho$, the switching rate function $\rho_{s s'}:\R^S\to \R_+$  of each pair of strategies $s,s'\in\cS$ is Lipschitz continuous:\footnote{\label{ftnt:L1norm}We adopt the $L^1$- norm as a norm on a finite-dimensional real space, which we denote by $|\cdot|$: for vector $\v=(v_i)_{i=1}^I\in\R^I$, $|\v|:=\sum_{i=1}^I |v_i|$.} there exists $L_\rho>0$ such that 
$$|\rho_{s s'}(\bpi)-\rho_{s s'}(\bpi')|\le L_\rho |\bpi-\bpi' | \qquad\text{ for any $s,s'\in\cS, \bpi,\bpi'\in\R^S$}.$$
\end{dfn}

\begin{exmp}
In a class of \textbf{pairwise comparison dynamics}, the switching rate $\rho_{s s'}(\bpi)$ increases with the payoff difference $\pi_{s'}-\pi_s$. In particular, the revision protocol $\rho_{s s'}(\bpi)=[\pi_{s'}-\pi_s]_+$ defines the \textbf{Smith dynamic} \citep{Smith84TranspSci_Stbl_Dyn_TrafficAssignmt}.\footnote{$[\cdot]_+$ is an operator to truncate the negative part of a number: i.e., $[\breve\pi]_+$ is $\breve\pi$ if $\breve\pi\ge 0$ and $0$ otherwise.}
\end{exmp}

\begin{exmp}\label{exmp:smoothBRD}
Because of continuity of a switching rate function, we see \textbf{smooth best response dynamics} \citep{FudenbergKreps93GEB_LearnMixed} as constructed from continuous revision protocols. For example, the \textbf{logit dynamic} \citep{FudenLevine98_Learinng} is constructed from $\rho_{s s'}(\bpi)=\exp(\mu^{-1}\pi_{s'})$ $ / \sum_{s''\in\cS}\exp(\mu^{-1}\pi_{s''} )$ with noise level $\mu>0$. \

This revision protocol can be obtained from perturbed optimization: upon the receipt of each revision opportunity, an agent draws each random perturbation in each strategy $s$'s payoff $\varepsilon_s$ from a double exponential distribution\footnote{Given the noise level $\mu$, the cumulative distribution function of the double exponential distribution is $\P(\varepsilon_s\le c)=\exp(-\exp(-\mu^{-1}c-\gamma))$ where $\gamma\approx 0.5772$ is Euler's constant.} and then switches to the strategy that maximizes $\pi_s +\varepsilon_s$ among all strategies $s\in\cS$. In general, a smooth best response dynamic can be constructed from such perturbed optimization under some admissibility condition: see \cite{HofbauerSandholm02EMA_StochFictPlay,HofSand07JET_RdmnDistPayoff}. Note that, after the receipt of a revision opportunity and a draw of $\boldsymbol{\varepsilon}\in\R^S$, an agent \textit{always }switches to the optimal strategy, however small the payoff gain by  this switch is.
\end{exmp}

Note that payoff perturbation $\boldsymbol{\varepsilon}=(\varepsilon_s)_{s\in\cS}$ is \textit{transient }: a different value of $\boldsymbol{\varepsilon}$ will be drawn at each revision opportunity from an i.i.d.~distribution. So, there is no (ex ante) heterogeneity in $\boldsymbol{\varepsilon}$. In contrast, the idiosyncratic payoff type $\btheta$ in our heterogeneous dynamics is \textit{persistent}.

\paragraph*{Exact optimization protocols.}  In an exact optimization protocol, an agent switches \textit{only }to the best response given the current payoff vector: if strategy $s'$ does not yield the maximal payoff among $\bpi =(\pi_1 ,\ldots,\pi_S )$, then $\rho_{s s'}(\bpi )=0$ regardless of the agent's current strategy $s$. We allow the switching rate to an optimal strategy to vary with $\bpi $ and $s,s'\in\cS$. Denote by $Q_{s s'}(\bpi )$ the \textit{conditional} switching rate from $s$ to $s'$, \textit{provided} that $s'$ is already designated as the new strategy. In the definition below, we extend the domain of $Q_{s s'}$ to $\R^S$ while assuming its continuity over the whole domain. The actual switching rate $\rho_{s s'}$ is defined as the truncation of $Q_{s s'}$ when $s'$ is not a best response; the truncation causes discontinuity.

\begin{dfn}[Exact optimization protocols]\label{dfn:ExactOpt}
In an \textbf{exact optimization protocol}, the switching rate function $\rho_{s s'}:\R^S\to \R_+$ of each pair of strategies $s,s'\in\cS$ is expressed as
$$ \rho_{s s'}(\bpi ) 
= \begin{cases} 
	0& \text{ if }s'\notin\argmax_{s''\in\cS} \pi_{s''} ,\\
	Q_{s s'}(\bpi ) &\text{ if } \{s'\}=\argmax_{s''\in\cS} \pi_{s''} ,
  \end{cases}$$
with a Lipschitz continuous function $Q_{s s'}:\R^S\to\R_+$.
\end{dfn}

\begin{exmp}
In the \textbf{standard best response dynamic (BRD)} as defined by  \cite{Hofbauer95BRD,GilboaMatsui91EMA_SocStbl}, a revising agent always switches to the optimal strategy that maximizes the current payoff with probability 1, however small the payoff gain by this optimization is. That is, the standard BRD is constructed from an exact optimization dynamic with $Q_{s s'}\equiv 1$. The heterogeneous version is considered in \cite{ElySandholmBayes1}; they prove that the aggregate strategy distribution in the heterogeneous standard BRD follows a  homogenized smooth BRD.
\end{exmp} 

\begin{exmp}
Consider a version of BRD in which the switching rate to the unique best response $Q_{s s'}$ depends on the payoff difference (the \textbf{payoff deficit}) between the current strategy $s$ and the best response $s'$, i.e., $Q_{s s'}(\bpi )=Q(\pi_{s'} -\pi_s )$ whenever $s'\in\argmax_{s''\in\cS}\pi_{s''} $. Function $Q:\R_+\to[0,1]$ is called a \textit{tempering function} and assumed to be continuously differentiable and satisfy $Q(0)=0$ and $Q(q)>0$ whenever $q>0$. Then this revision protocol yields the \textbf{tempered BRD}; \cite{ZusaiTBRD} constructs this revision protocol from optimization with a stochastic switching cost whose cumulative distribution function is $Q$. 
\end{exmp} 

\section{Existence of a unique solution trajectory}\label{sec:Exist}
We verify Lipschitz continuity of a heterogeneous dynamic to guarantee the existence of a unique solution trajectory from an arbitrary initial strategy distribution. We use a version of Picard-Lindel\:{o}f theorem (\Cref{thm:Zeidler_SolExist} in \Cref{apdx:Lcont_dyn}) to obtain the existence of a unique solution trajectory from Lipschitz continuity of heterogeneous dynamic $\V^\F$. To apply this theorem, the domain of the dynamic must be a Banach (complete normed vector) space. Since the set of strategy distributions is not a vector space, we extend the domain to the space of finite signed measures $\cM_{\cS\Theta}$ with a variational norm $\|\cdot\|^\infty_{\cS\Theta}$ (see \Cref{apdx:norm}) and then prove the Lipschitz continuity of $\V^\F$ on this extended domain as in \eqref{eq:Lcont_V} in \Cref{apdx:Lcont_dyn}. 

For this, we assume that payoff function $\F$ is Lipschitz continuous and switching rate function $\rho_{\cdot\cdot}$ is bounded. It may sound too restrictive if we require switching rate to be bounded while extending the domain to $\cM_{A\times\Theta}$, which might be interpreted as allowing the population size to be any finite mass. However, as we will truncate a finite signed measure $\M\in\cM_{\cS\Theta}$ by rounding off too high densities in our proof (see \Cref{apdx:Lcont_dyn}), it is enough to require those regularity conditions to hold a subset of $\cM_{\cS\Theta}$: let $\bar\cM_{\cS\Theta}\subset\cM_{\cS\Theta}$ be the set of finite signed measures that are absolutely continuous w.r.t.~$\P_\Theta$ and whose density $\m$ is \textit{truncated} by $\pm 3$, i.e., $m_s(\btheta)\in[-3,3]$ for any $s\in\cS$ and $\P_\Theta$-almost all $\btheta$.\footnote{The bound $\pm 3$ can be replaced with any fixed numbers; here it is chosen just to be consistent with our (arbitrary chosen) bound on the rounding function; see \Cref{apdx:Lcont_dyn}. Note that the density $\x$ of $\X\in\cX$ is bounded as $\x(\btheta)\in\Delta^\cS$ and thus $x_s(\btheta)\in[0,1]$; thus, we do not truncate joint strategy distribution $\X\in\cX$ and so $\cX\subset\bar\cM_{\cS\Theta}$.}

\begin{assmp}[Lipschitz continuity of the payoff function]\label{ass:F0}
For $\P_\Theta$-almost every type $\btheta\in\Theta$, $\F(\btheta):\R^S\to\R^S$ is Lipschitz continuous with Lipschitz constant $L_\F(\btheta)$:
$$ | \F[\M](\btheta) - \F[\M'](\btheta) | \le L_\F(\btheta) \| \M-\M' \|^\infty_{\cS\Theta} \qquad \text{ for any $\M,\M'\in\bar\cM_{\cS\Theta}$}.$$
In addition, $\bar L_\F:=\E_\Theta L_\F<\infty$.\footnote{\Cref{ass:F0} is satisfied in an ASAG, as long as the common payoff function $\F^0:\R^S\to\R^S$ is Lipschitz continuous.}
\end{assmp}


\begin{assmp}[Bounded switching rates]\label{ass:bddR}
There exists $\bar\rho\in\R_+$ such that 
$$ \rho_{s s'}(\F[\M](\btheta))\le \bar\rho \quad\text{ for any $s,s'\in\cS$ and $\P_\Theta$-almost all $\btheta\in\Theta$}$$
for any truncated finite signed measure $\M\in\bar\cM_{\cS\Theta}$.\footnote{\Cref{ass:bddR} is satisfied in an ASAG, if the type distribution $\P_\Theta$ has a bounded support and the common payoff function $\F^0$ is continuous, even if the switching rate function itself is not bounded over the whole domain $\R^S$ like the Smith dynamic.} 
\end{assmp}

For an exact optimization protocol, the Lipschitz continuity of $Q_{s s'}$ is not sufficient to guarantee Lipschitz continuity of revision protocol $\brho:\R^S\to\R^{S\times S}$ due to truncation when the best response strategy changes. The continuity of $Q_{s s'}$ assures continuous change in switching rate $\rho_{s s'}$ with the payoff vector, when strategy $s'$ remains to be the unique best response. However, payoff changes may cause changes in the best responses, which triggers discontinuous changes in the switching rates: the switching rate $\rho_{s s'}$ to the new best response strategy changes from zero to some positive rate $Q_{s s'}$ and the switching rate to the old one changes from positive to zero. The next assumption states that, when the joint strategy distribution changes, the mass of agents who experience switches of best responses grows only continuously with the distance between old and new joint strategy distributions; thus, despite discontinuous changes in individual agents' switching rates, the sum of these changes over all the agents is continuous.

\begin{assmp}[Continuous change in best response]\label{ass:cont_type}
\textit{If} revision protocol $\brho:\R^S \to \R^{S\times S}_+$ is an exact optimization protocol, then there exists $L_\BR\in\R_+$ such that 
$$ \P_\Theta(\Theta^\F_{s\in\BR}(\M) \cap \Theta^\F_{s'\in\BR}(\M'))\le L_\BR \| \M-\M' \|$$
for any two distinct strategies $s,s'\in\cS$ such that $s\ne s'$ and any two truncated finite signed measures $\M,\M'\in\bar\cM_{\cS\Theta}$.\footnote{In an ASAG, \Cref{ass:cont_type} is satisfied if the distribution of differences in idiosyncratic payoffs between every two strategies satisfies a Lipschitz-like continuity in the sense that there exists $\bar p_\Theta\in\R$ such that
$\P_\Theta(\{\btheta\in\Theta : c\le \theta_{s'}-\theta_s\le d\})\le (d-c) \bar p_\Theta  $
for any $s,s'\in\cS$ and any $c, d\in \R$ such that $d>c$.}
\end{assmp}

This assumption implies that the best response is unique for $\P_\Theta$-almost all types (let $\M=\M'$). Note that this assumption imposes the condition on the type distribution only if the revision protocol is an exact optimization protocol; L-continuous revision protocols do not need any such assumption on the type distribution for the existence of a unique solution trajectory.


\begin{thm}[Lipschitz continuity of $\V^\F$]\label{thm:Lcont_dyn}
Consider a heterogeneous dynamic $\V^\F$ in a population game $\F$ under an L-continuous revision protocol or an exact optimization protocol. Under \Cref{ass:F0,ass:bddR,ass:cont_type}, function $\V^\F$ is Lipschitz continuous in variational-norm $\|\cdot\|^\infty_{\cS\Theta}$ over $\cM_{\cS\Theta}$.
\end{thm}

\begin{cor}[Existence of a unique solution trajectory]\label{thm:ExistTraj_dyn}
If $\V^\F$ satisfies the assumptions for \Cref{thm:Lcont_dyn}, then there exists a unique solution trajectory $\{\X_t\}_{t\in\R_+}\subset \cX$ of $\dot\X_t=\V^\F[\X]$ from any initial  strategy distribution $\X_0\in\cX$.
\end{cor}
%

Appendix \ref{apdx:model} explains the whole outline of the proof as well as the measure-theoretic construction of heterogeneous dynamics, while heavy calculation in the proof is postponed to Section \ref{supp:Model} of  Supplementary Note. The basic idea of the proof follows a conventional proof for continuous-\textit{strategy} dynamics such as in \cite{OechsslerRiedelET01_InfStrEvolDyn} and \cite{Cheung13_PairwiseCompDyn_ContStr}. Both ours and theirs deal with the dynamic of probability measure on a (possibly) continuous space. However, there are two major differences in the proof for our  continuous-\textit{type} dynamics, compared to that for their continuous-\textit{strategy} dynamics, as discussed in the remarks below. 

\begin{rmk}\label{rmk:Lcont_NeedDensity} 
One of the differences comes from the essential defining nature of heterogeneous dynamics that each agent is born with a certain type $\btheta$ and posses it persistently; so $X_s(B_\Theta)$ can never exceed $\P_\Theta(B_\Theta)$ for any $B_\Theta\in\cB_\Theta, s\in\cS$. As argued in \Cref{sec:Model}, this assures $\P_\Theta\gg \X$, i.e., the absolute continuity of $\X$ with respect to $\P_\Theta$. This enables us to obtain a conditional strategy distribution $\x$ as a density of $\X$ w.r.t.~$\P_\Theta$ and to interpret $x_s(\btheta)$ as a proportion of strategy-$s$ players among type-$\btheta$ agents. Since an agent's strategy revision crucially depends on the own type, it is natural to construct dynamic $\v$ of conditional strategy distribution $\x(\btheta)$ at each $\btheta$, as in \eqref{eq:Dyn_x}; then, dynamic $\V$ of strategy distribution $\X$ is just derived from $\v$, as in \eqref{eq:Dyn_composite} in \Cref{apdx:Lcont_dyn}. 

When extending the domain of $\V$ to $\cM_{\cS\Theta}$, a finite signed measure may not be absolute continuous with respect to $\P_\Theta$ and thus may not have a density. Then, how can we extend the density-based definition \eqref{eq:Dyn_x} of our dynamic? For this, we extract the absolute continuous part of a measure by Lebesgue decomposition theorem (\Cref{lem:LebesgueDecomp}) to keep the density-based construction of the dynamic. 

On the other hand, in continuous-\textit{strategy} evolutionary dynamics, an agent is assumed to be homogeneous and thus has no persistent  characteristic. When they need some distribution that dominates the strategy distribution to obtain absolute continuity, they \textit{create }some \textit{ad hoc }distribution artificially from the strategy distribution.\footnote{For example, see \citet[p.159]{OechsslerRiedelET01_InfStrEvolDyn} and \citet[p.2 in Online Appendix]{Cheung13_PairwiseCompDyn_ContStr}.} A continuous-strategy evolutionary dynamic typically defines the transition of the measure (the mass of players in a Borel \textit{set }of strategies) directly; they obtain a density only to prove Lipschitz continuity. Thus, a dominating distribution for absolute continuity is only an artificial addition to continuous-\textit{strategy} evolutionary dynamics, not an essential component of games or dynamics.
\end{rmk}

\begin{rmk}\label{rmk:Lcont_DiscontExactOpt} 
Another difference is that we cover exact optimization dynamics such as the standard and tempered BRDs, whose revision protocols $\brho$ are  discontinuous. As far as the author is aware of, the studies on continuous-strategy evolutionary dynamics focus on L-continuous revision protocols:  imitative dynamics \citep{OechsslerRiedelET01_InfStrEvolDyn,Cheung16GEB_ImitatDyn_ContStr}, the BNN dynamic \citep{HofOechsslerRiedel09GEB_BvNDyn_ContStr}, the gradient dynamic \citep{FriedmanOstrov13JET_EvolDyn_ContsActions}, payoff comparison dynamics \citep{Cheung13_PairwiseCompDyn_ContStr} and the logit dynamic \citep{LahkarRiedel15GEB_LogitDyn_ContStr}. 

In our heterogeneous exact optimization dynamics, we suppress discontinuity in switching rates by continuity in the mass of types of agents who experience discontinuous changes in switching rates, as assumed in \Cref{ass:cont_type}. This continuity of the type distribution mitigates discontinuity in switching rates and helps to retain continuity of the dynamic, thanks to $\P_\Theta\gg\X$. 

For continuous-\textit{strategy} dynamics, one might assume continuity of the \textit{ad hoc }dominating distribution, argued in the above remark. But it restricts the strategy distribution to being continuous on the strategy space. However, since agents are homogeneous in these dynamics, it would be commonly expected that every agent eventually takes the same strategy, when the game has a unique pure-strategy Nash equilibrium. For this, the assumption of continuity in the strategy distribution is too demanding.
\end{rmk}

\section{Equilibrium stationarity and stability}\label{sec:EqmComp_Ext}
Our heterogeneous dynamics could be seen as an extension of evolutionary dynamics in a single homogeneous population to (possibly) continuously many heterogeneous subpopulations, though the existence of a unique solution trajectory requires careful formulation of the state space. It is natural to expect that stationarity and stability of Nash equilibria are extended to \textit{equilibrium strategy distributions }in the heterogeneous setting. 

We first define the properties of mean dynamic $\v$ that induce stationarity and stability of equilibria, separately from the population game.\footnote{This separation accords with the view proposed by \cite{SandholmPopText} and \cite{FoxShamma13Games}. \citet[especially, Sec. 1.2.2 and Ch.4]{SandholmPopText} proposes to construct a mean dynamic $\v$ from agents' revision protocol $\brho$ and thus has guided our attention to individual decision rules behind the collective population dynamic. Pushing further in this direction, \cite{FoxShamma13Games} regard an evolutionary dynamic $\v^\F$ as a hybrid of a mean dynamic $\v$ (a system that converts payoff vector $\bpi$ to the transition of state $\dot\x$) and a population game $\F$ (a feedback system that converts the current state $\x$ to payoff vector $\bpi$); they propose to define properties of $\v$ (e.g. PC in our paper; $\delta$-passivity in their paper) apart from those of $\v^\F$ (e.g. equilibrium stability of potential games in our paper, and of contractive games in their paper), which are derived by combining properties of $\v$ and assumptions on $\F$.} This separation is useful because both homogeneous and heterogeneous dynamics stem from the same mean dynamic $\v$ (constructed from the same revision protocol $\brho$). Their difference lies only in the difference in the population game played by agents, namely the difference between $\F:\cX\to\cC_\Theta$ and $\F^0:\Delta^S\to\R^S$.

In the homogeneous setting, the stationarity of a Nash equilibrium under $\v^\F$ is an immediate consequence of the \textit{best response stationarity} under $\v$; the mean dynamic stays at a strategy distribution if and only if agents are taking the best response to the current payoffs.
\begin{dfn}[Best response stationarity of mean dynamic]\label{dfn:BRStat}
Mean dynamic $\v:\Delta^S\times\R^S\to\R^S$ satisfies the \textbf{best response (BR) stationarity} if, for any $\bpi^0\in\R^S, \x^0\in\Delta^S$,
\begin{equation}
\v(\bpi^0,\x^0)=\0 \qquad \Longleftrightarrow \quad 
\x^0\in\Delta(\cS_\BR(\bpi^0)). \label{eq:BRS_type}
\end{equation}
\end{dfn}

All the evolutionary dynamics mentioned in Section \ref{sec:dyn}, except smooth BRDs, satisfy BR stationarity.\footnote{In the homogeneous version of exact optimization dynamics, the best response stationarity needs to assume $\rho_{s s'}(\bpi)=0$ when the current strategy $s$ is a best response to $\bpi$; this was not assumed in our definition in cases of multiple best responses. In the heterogeneous setting, this concern on multiple best responses is eliminated by \Cref{ass:cont_type}. Hence, this assumption replaces the assumption of $\rho_{s s'}(\bpi)=0$ for best response $s$ to $\bpi$.} In a homogeneous population game, the best response stationarity implies the stationarity of a Nash equilibrium and non-stationarity of non-equilibrium states.

The key property of evolutionary dynamics for equilibrium stability  is the \textit{positive correlation}: each strategy's payoff  and the net increase in the mass of the strategy's players are positively correlated and the correlation is strictly positive unless the strategy distribution is unchanged. Major evolutionary dynamics, except smooth BRDs, satisfy the positive correlation.

\begin{dfn}[Positive correlation of mean dynamic]
\label{dfn:PC}
Mean dynamic $\v:\Delta^S\times\R^S\to\R^S$ satisfies the \textbf{positive correlation (PC)} if 
\begin{equation}
\bpi^0\cdot\v(\bpi^0,\x^0) 
	\begin{cases}
	\ge 0 & \text{ for any }\bpi^0\in\R^S, \x^0\in\Delta^S; \\ 
	>0 &\text{ if }  \v(\bpi^0,\x^0)\ne \0.
	\end{cases} \label{eq:PC_type}
\end{equation}
\end{dfn}

While stability of a Nash equilibrium is not generally guaranteed even in the homogeneous setting, it is assured for potential games under a wide class of evolutionary dynamics. In the homogeneous setting, population game $\F^0:\Delta^\cS\to\R^S$ is a potential game if there is a differential function $f^0:\Delta^\cS\to\R$, called a potential function, such that $\nabla f^0\equiv\F^0$.\footnote{Having a potential function is equivalent to externality symmetry: the change in the payoff of a strategy by a change in the mass of another strategy's players is symmetric between these two strategies. The class of potential games includes random matching in common interest games, binary games and congestion games. \citet[Chapter 3]{SandholmPopText} provides further explanation and examples.} PC immediately implies that the value of $f^0$ increases over time until it reaches a stationary point, since the definition of a potential function implies $ \dot{f}^0(\x^0)=\nabla f^0(\x^0)\cdot \dot\x^0=\F^0(\x^0)\cdot \dot\x^0= \F^0(\x^0)\cdot\v(\F^0(\x^0),\x^0).$
Thus, the homogeneous {\it potential} function $f^0$ works as a Lyapunov function commonly in these evolutionary dynamics and so PC assures stability of local maxima of $f^0$ \citep{Sandholm01JET_Potential}. 


If a dynamic satisfies BR stationarity and PC, we call it an \textbf{admissible} dynamic. Pairwise comparison dynamics and exact optimization dynamics are admissible dynamics.\footnote{Smooth BRDs satisfy analogous properties of BR stationarity and PC for perturbed payoffs; see \citet[\S 6.2.4]{SandholmPopText}. For observational dynamics such as the replicator dynamic and excess payoff dynamics, see \Cref{sec:Observ}. See \citet[Chapter 5]{SandholmPopText} for summary of the relationship between dynamics and the two properties in this section.}  

\subsection{Equilibrium Stationarity in general}
In the heterogeneous setting, the best response stationarity applies to each type: the strategy distribution of a particular type $\btheta$ remains unchanged if and only if almost all agents of this type choose the best response to the current payoff for this type. Thus, it is straightforward that the best response stationarity implies the stationarity of an equilibrium strategy distribution and non-stationarity of non-equilibrium strategy distributions.


\begin{thm}[Equilibrium stationarity]\label{thm:NStat}
Suppose that mean dynamic $\v$ satisfies the best response stationarity \eqref{eq:BRS_type}. %
Then, in any heterogeneous population game $\F$, an equilibrium strategy distribution is stationary under the heterogeneous evolutionary dynamic $\V^\F$ derived from these $\v$ and $\F$, and vice versa:\footnote{$\O=(O_s)_{s\in\cS}\in\cM_{\cS\Theta}$ denotes a zero measure such as $O_s(B_\Theta)=0$ for any $B_\Theta\in\cB_\Theta, s\in\cS$.}
\begin{equation}
\V^\F[\X]=\O  \qquad \Longleftrightarrow \qquad \X\text{ is an equilibrium strategy distribution in }\F. \label{eq:NS_comp}
\end{equation} 
\end{thm}

This theorem suggests that the existence of a stationary point is equivalent to that of an equilibrium state. Following the outline of the proof for the existence of a distributional equilibrium in an incomplete information game of finitely many players by \cite{MilgromWeber85MathOR_DistrStr}, we can guarantee the existence of an equilibrium strategy distribution in a heterogeneous population game that exhibits a kind of uniform continuity and boundedness over types of the payoff function.\footnote{Just like \cite{MilgromWeber85MathOR_DistrStr},  we use Glicksberg's fixed point theorem \cite[henceforth AP; Corollary 17.55]{AliprantisBorder_InfiniteDimAnal}, since an equilibrium strategy distribution can be formulated as a fixed point of the ``distributional'' best response correspondence (check $B[\X]$ in \Cref{apdx:ExistEqm}). But the objective function in the best response correspondence is different from theirs; so we need to prove continuity of the objective function specifically for our setting.}

\begin{thm}[Existence of equilibrium]\label{thm:ExistEqm}
Suppose that $\F:\cX\to\cC_\Theta$ satisfies \Cref{ass:F0},  \textbf{equicontinuity over types} (with respect to the weak topology metrized by Prokorov metric $d_\cM$),\footnote{See \Cref{sec:comp_Bayes} for the weak topology and Prokorov metric $d_\cM$.} i.e., for each $\X\in\cX$ and for any $\varepsilon>0$, there exists $\delta_\text{Ct}[\X]>0$ such that
$$ d_\cM(\X,\X') <\delta_\text{Ct}[\X] \quad\Longrightarrow\quad \left[ | F_s[\X](\btheta)-F_s[\X'](\btheta) | <\varepsilon \quad\text{for any $s\in\cS$ and $\P_\Theta$-almost all $\btheta\in\Theta$}\right],$$
and \textbf{near-boundedness over types}, i.e., for each $\X\in\cX$ and for any $\varepsilon>0$, there exists a combination of $\bar F[\X]\ge 0$ and $\delta_\text{Bd}[\X]>0$ such that for any $\Y,\Y'\in\cX$\footnote{Here $[\cdot]_+$ is an operator such as $[z]_+=\max\{0,z\}$. If $\F[\X]:\Theta\to\R^S$ is bounded over $\Theta$, i.e., there exists $\bar F[\X]$ such that $|F_s[\X](\btheta)| \le \bar F[\X]$ for any $s\in\cS$ and $\P_\Theta$-almost any type $\btheta$, then it is nearly bounded over types.} 
\begin{align*}
& d_\cM(\Y,\Y')<\delta_\text{Bd}[\X] \quad\Longrightarrow\quad \left| \int_\Theta \sum_{s\in\cS} \left[ |F_s[\X](\btheta)| -\bar F[\X] \right]_+ \cdot (Y_s(d\btheta)-Y'_s(d\btheta)) \right| <\varepsilon.
\end{align*} 
Then, there exists an equilibrium strategy distribution in the heterogeneous population game $\F$. 
\end{thm}

\begin{cor} Under the assumptions for \Cref{thm:NStat,thm:ExistEqm}, heterogeneous dynamic $\V^\F$ has a stationary state.
\end{cor}



\subsection{Equilibrium stability in potential games}
\subsubsection*{Heterogeneous potential games}
For a game played in large population, a potential game is defined as a game in which payoff vector can be derived as the derivative of some scalar-valued function, i.e., a potential function. That is, the payoff vector is a coefficient vector in linear approximation of the change in the potential value. By generalizing this idea to a function defined on the (possibly infinite-dimensional) space of strategy distributions, we define a heterogeneous potential game.


\begin{dfn}[Heterogeneous potential game]\label{dfn:Pot_Hetero}
Heterogeneous population game $\F:\cS\to\cC_\Theta$ is called a {\bf heterogeneous potential game} if there is a scalar-valued Fr\'{e}chet-differentiable function $f:\cX\to\R$ that is continuous in the weak topology on $\cX$ and whose Fr\'{e}chet derivative coincides with the payoff function: at each strategy distribution $\X\in\cX$, the payoff vector function $\F[\X]\in\cC_\Theta$ satisfies\footnote{Here, operator $\langle\cdot,\cdot\rangle$ is defined as $\langle \bpi,\Delta\X \rangle=\int_\Theta \sum_{s\in\cS} \pi_s(\btheta) \Delta X_s(d\btheta)$ and $\Delta\X \in\cM_{\cS\Theta}$. The norm $\|\cdot\|^\infty_{\cS\Theta}$ is the variational norm on $\cX$ to metrize the strong topology: we have $\|\Delta\X\|^\infty_{\cS\Theta}=\sum_{s\in\cS} \E_\Theta|\Delta x_s|$ by \Cref{thm:VarNorm} in Appendix  \ref{apdx:norm}. Fr\'{e}chet differentiability is defined for the strong topology and thus continuity in the weak topology is additionally required.}
$$ f(\X')=f(\X)+\langle \F[\X], \X'-\X \rangle +o(\|\X'-\X\|^\infty_{\cS\Theta}) \qquad\text{for any $\X'\in\cX.$}$$
We call $f$ a (heterogeneous) potential function for $\F$.
\end{dfn}

Naturally in our canonical three examples, if the base game is a potential game, then its heterogeneous versions are also potential games.


\begin{exmppr}{exmp:ASAG_Dfn}
Recall \Cref{exmp:ASAG_Dfn}. Assume that the base homogeneous game $\F^0$ is a (homogeneous) potential game with potential function $f^0:\Delta^\cS\to\R$ such that $\nabla f^0\equiv \F^0$. Then, the ASAG is a heterogeneous potential game, with potential function $f:\cX\to\R$ such as\footnote{This function $f$ appears in the study of evolutionary implementation by \citet[Appendix A.3]{SandholmNegImpl}. But it was used there only to characterize an equilibrium as a solution of the KKT condition for local maxima and minima of $f$.}
\begin{equation}
f(\X)= f^0(\X(\Theta))+\int_\Theta \btheta\cdot \X(d\btheta)  \qquad\text{ for each }\X\in\cX.\label{eq:Lyap_pot}
\end{equation}
\end{exmppr}

\begin{exmppr}{exmp:RndMat_IncompInfo}
Recall \Cref{exmp:RndMat_IncompInfo}. Now assume the base two-player game $U$ is a potential game. For a two-player game, it means that the payoff $U(\theta,\theta')$ is decomposed as $U(\theta,\theta')=U^0(\theta,\theta')+\1 \vec r(\theta,\theta')$ with a symmetric matrix $U^0(\theta,\theta')$ and a row vector $\vec r(\theta,\theta')$. Further, we assume that $U^0(\theta,\theta')=U^0(\theta',\theta)$. Hence, $u_{ss'}(\theta,\theta')$ is decomposed to $u_{ss'}^0(\theta,\theta')+r_{s'}(\theta,\theta')$, where the former term represents the common payoff for the two matched agents such that $u_{ss'}^0(\theta,\theta')=u_{a'a}^0(\theta,\theta')=u_{a'a}^0(\theta',\theta)$ and the latter term represents the payoff that is independent of the agent's own strategy $s$. Then, the random-matching population game $\F$ is a potential game, with potential function $f:\cX\to\R$ such that
$$ f(\X)=0.5\int_{(\theta,\theta')\in\Theta^2} \X(d\theta)\cdot U^0(\theta,\theta') \X(d\theta') \qquad\text{ for each }\X\in\cX.$$
\end{exmppr}

\begin{exmppr}{exmp:RndMat_TransitSig}
Recall \Cref{exmp:RndMat_TransitSig}. Now assume the base two-player game $U$ is a potential game,\footnote{See \cite{Ui09IJET_BayesPot_InfoStr_Team} for examples of Bayesian potential games.} and thus the payoff $U(\theta,\theta')$ is decomposed as $U(\theta,\theta';\hat\theta,\hat\theta')=U^0(\theta,\theta';\hat\theta,\hat\theta')+\1 \vec r(\theta,\theta';\hat\theta,\hat\theta')$ with a symmetric matrix $U^0(\theta,\theta';s,s')$ and a row vector $\vec r(\theta,\theta';\hat\theta,\hat\theta')$. Further, we assume that $U^0(\theta,\theta';\hat\theta,\hat\theta')=U^0(\theta',\theta;\hat\theta',\hat\theta)$ and $\P_{\hat\Theta}(\hat\theta,\hat\theta'|\theta,\theta')=\P_{\hat\Theta}(\hat\theta',\hat\theta|\theta',\theta)$. 

Then, the random-matching population game is a potential game, with potential function $f:\cX\to\R$ such that
$$ f(\X)=0.5\int_{(\theta,\theta')\in\Theta^2} \sum_{s'\in\cS}  \sum_{(\theta,\theta')\in \hat\Theta^2} u_{s(\hat\theta) s'(\hat\theta')}(\theta,\theta';\hat\theta,\hat\theta') \P_{\hat\Theta}(\hat\theta,\hat\theta'|\theta,\theta') X_{s'}(d\theta') X_s(d\theta) \qquad\text{ for each }\X\in\cX.$$
\end{exmppr}

\begin{exmppr}{exmp:StrPop}
Recall \Cref{exmp:StrPop}. Now assume that the base game is a potential game with potential function $f^0:\Delta^\cS\times\Delta^\cS\to\R$, i.e., $\nabla_1 f^0(\x,\x')=\F^0(\x,\x'), \nabla_2 f^0(\x,\x')=\F^0(\x',\x)$, where $\nabla_i f^0$ is the gradient vector of $f^0$ with respect to the strategy distribution in the $i$-th argument. (Recall the first argument is the strategy distribution in the own population and the second is that in the opponent.) Further, assume that the weight function $g$ is symmetric: $g(\theta,\theta')=g(\theta',\theta)$. Then, the structured population game is a potential game, with potential function $f:\cX\to\R$ such that
$$ f(\x)=0.5\int_{(\theta,\theta')\in\Theta^2} f^0(\x(\theta),\x(\theta')) g(\theta,\theta') \P_\Theta(d\theta)\P_\Theta(d\theta') \qquad\text{ for each }\X\in\cX.$$
\end{exmppr}

\begin{thm}\label{thm:potExmp}
The heterogeneous population games in the above examples are indeed heterogeneous potential games.
\end{thm}

Both in the homogeneous and heterogeneous settings, all local maxima and interior local minima of a potential function, and indeed all the solutions of the Karash-Kuhn-Tucker first-order condition for maxima are equilibria in a potential game; see \cite{Sandholm01JET_Potential} for the proof for Nash equilibria in a homogeneous potential game and \citet[Appendix A.3]{SandholmNegImpl} for equilibrium strategy distributions in a heterogeneous potential game.

\subsubsection*{Stability and potential maximization}
In the heterogeneous setting, the positive correlation of $\v$ implies a positive correlation between the payoffs and the strategy distribution among \textit{each type} of agents. Thus, by the same token as in a homogeneous potential game, this guarantees that the heterogeneous potential function $f$ works as a Lyapunov function for equilibrium stability in a heterogeneous potential game $\F$. 

\begin{thm}[Equilibrium stability of heterogeneous potential games]\label{thm:NStbl_pot}
Suppose that mean dynamic $\v$ satisfies the positive correlation \eqref{eq:PC_type} as well as \Cref{ass:F0,ass:bddR,ass:cont_type}. Then, in any heterogeneous potential game $\F$, the following holds. 
\begin{enumerate}[i)]
\item The set of stationary strategy distributions $\{\X\in\cX : \V^\F[\X]=\O\}$ is globally attracting under $\V^\F$. A local maximum (local strict maximum, resp.) of $f$ is Lyapunov stable (asymptotically stable, resp.). 
\item Let $\X^\ast$ be an isolated stationary strategy distribution in the sense that, in a neighborhood $\cX^\ast$ of $\X^\ast$ in the space $\cX$, there is no other stationary strategy distribution than $\X^\ast$. a) If it is (locally) asymptotically stable, then it is a local strict maximum of $f$. b) Further assume that $\gamma:\cX\to\R$ defined as $\gamma(\x)=\langle \F[\X], \V^\F[\X] \rangle$ is continuous in weak topology.\footnote{This is continuous in strong topology if $\F$ and $\V^\F$ are continuous in strong topology, which is guaranteed by \Cref{ass:F0} and \Cref{thm:Lcont_dyn}. But continuity in weak topology is stronger than that in strong topology, since convergence in the former is weaker than that in the latter (and then the value of a ``continuous'' function must approach arbitrarily close to a limit).} If $\X^\ast$ is Lyapunov stable, then it is a local maximum of $f$. 
\end{enumerate}
\end{thm}

As a local maximum of $f$ is an equilibrium strategy distribution and BR stationarity implies equilibrium stationarity, this theorem suggests global convergence to the set of equilibrium strategy distributions under admissible dynamics. Furthermore, since equilibria and potential maximizers can be found solely from $\F$ independently of the dynamic $\v$, \Cref{thm:NStbl_pot,thm:NStat} suggest that the set of stationary states and the set of locally stable states are common to all admissible dynamics. 

\begin{cor}\label{cor:NStbl_pot_SpecifDyn}
Consider admissible dynamics in a heterogeneous population game $\F$ that satisfies \Cref{ass:F0,ass:bddR,ass:cont_type}.\footnote{In each of the two claims i,ii), the former condition (equilibrium/potential maximum) is sufficient for the latter (stationarity/stability) to hold under \textit{all} admissible heterogeneous dynamic, while the latter under \textit{any }(single) admissible dynamic is sufficient for the former.}

\begin{enumerate}[i)]
\item Strategy distribution $\X^\ast$ is an equilibrium in $\F$, if and only if it is stationary in any admissible dynamic $\V^\F$. 
\item Further, suppose that $\F$ is a heterogeneous potential game. Then, the set of equilibrium strategy distributions is globally attracting. Isolated equilibrium strategy distribution $\X^\ast$ is a local strict maximum of potential function $f$, if and only if it is asymptotically stable in any admissible dynamic $\V^\F$.
\end{enumerate}
\end{cor}

If $\X^\ast$ attains the global strict maximum, then it must be globally asymptotically stable in any admissible dynamics. Thus, once we establish global asymptotic stability of $\X^\ast$ under some particular admissible dynamic in a potential game, then it can be carried to all admissible dynamics. We see below applications of this positive result to ASAGs. 

\begin{exmp}[Convergence to a free-entry equilibrium.]\label{exmp:Entry_NegExt}
Consider a binary homogeneous game $\cS=\{I,O\}$ with \textit{negative} externality: $F^0_I(\bar x_I)$ decreases with $\bar x_I\in[0,1].$ Then, the potential function $f^0:[0,1]\to\R$ is given by $f^0(\bar x_I)=\int^{\bar x_I}_0 F^0_I(\bar y) d\bar y $ and strictly concave. With the boundedness of the domain $[0,1]$, the strict concavity of $f^0$ implies that the global maximum exists uniquely and there is no other local maximum of $f^0$. The global maximum of $f^0$ is the only equilibrium of this game. 

For an example in microeconomic theory to fall into this class of games, consider an entry-exit game played by suppliers in a particular industry. To make entry and exit symmetric, it is conventionally assumed that fixed costs exist but they are not sunk: fixed costs are paid only to maintain production capacities and they are revocable when the supplier becomes inactive. Further, the choice of entry or exit is conventionally regarded as a ``long run'' decision while the choice of the quantity supplied is a  ``short run'' decision (as well as the underlying consumers' decisions on the demand side); thus, it is commonly assumed that the market is settled to a market equilibrium (the state where the demand equals to the total supply) at each moment of time, given the mass (number) of active suppliers at the moment. A free-entry or so-called ``long run'' equilibrium is characterized in the homogeneous setting as a state in which the gross profit for an active producer is equal to the fixed cost. 

One may want to introduce heterogeneity in the suppliers' fixed costs; it not only sounds realistic but also eliminates indeterminacy of individual choices at a free-entry equilibrium. Under the heterogeneity in fixed costs, a free-entry equilibrium should be redefined as a state in which almost all the active producers have smaller fixed costs than the gross profit and almost all the inactive ones have greater fixed costs.

Under perfect competition in a standard setting as in \citet[Section 10.F]{MWG95}, the instantaneous market-equilibrium profit of an active supplier decreases with the number of active suppliers. We can regard $F^0_I(\bar x_I)$ as the gross profit at this instantaneous competitive equilibrium given the current mass $\bar x_I$ of active suppliers and $\theta_O(\omega)$ as the fixed costs of supplier $\omega$, while setting $F^0_O\equiv 0$ and $\theta_I\equiv 0$ for all agents; then, the choice between entry and exit in perfect competition falls into an ASAG with negative externality.
 
Thanks to our stability result, we can justify the free-entry equilibrium as the globally stable state in an evolutionary dynamic; indeed it is so strengthened to be stable in any admissible dynamics. As argued in \cite{ZusaiTBRD}, the tempered BRD is considered as a version of the BRD in which a revising agent pays a stochastic switching cost. Thus, the stability of the free-entry equilibrium under the tempered BRD suggests in this context that, even if entry and exit incur sunk costs to build or scrap the production capacity, the ``long-run'' equilibrium is indeed the long-run limit state under such an entry-exit dynamic.

By the same token, we can justify a free-entry equilibrium in the standard (static) monopolistic competition model such as \cite{DixitStiglitz77AER_MonopolComp} as a dynamically stable state under an arbitrary admissible dynamic.
\end{exmp}

\begin{exmp}[Dynamic implementation of the social optimum.]
Imagine a central planner whose goal is to maximize the total payoff of agents in an ASAG: 
$$ \E_\Theta \left[ \F[\X](\btheta)\cdot\x(\btheta) \right]=\F^0(\bar\x)\cdot\bar\x +\E_\Theta \left[ \btheta \cdot\x(\btheta) \right] \quad\text{ with }\bar\x=\X(\Theta), \X=\int \x d\P_\Theta.$$
To help the central planner achieve this goal, we introduce a monetary transfer to the agent's payoff: now a type-$\btheta$ agent's payoff from strategy $s\in\cS$ is $\tilde F^{\mathbf T}_s[\x](\btheta):=F_s[\x](\btheta)-T_s[\bar\x]$, where function ${\mathbf T}=(T_s)_{s\in\cS}:\Delta^\cS\to\R^S$ is a pricing scheme to determine the amount of the monetary transfer (in terms of payoff) from the agent to the planner for taking each strategy given aggregate strategy distribution $\x\in\Delta^\cS.$ 

In the homogeneous setting, a desirable aggregate state could be achieved by a very simple bang-bang control that gives a subsidy for strategies that need more players and imposes a tax on strategies that need less. By keeping the taxes and subsidies at extreme levels in their feasible ranges, convergence can be achieved in a finite time; and it is the fastest among all the pricing schemes. But, in the heterogeneous setting, such extreme pricing may result in excessive distortion of the underlying strategy distribution and practically unacceptable instability. To avoid these troubles, pricing should be less extreme and adjusted continuously over time. 

\cite{SandholmCongstPr,SandholmNegImpl} proposes the dynamic Pigouvian pricing scheme such as
$$ T_s[\bar\x] = -\sum_{s'\in\cS} \bar x_{s'} \pdif{F^0_{s'}}{\bar x_s}(\bar\x)  \quad\text{ for each }\bar\x\in \Delta^S.$$
Notice that this pricing scheme does not require the central planner to know  agents' revision protocols, the type distribution, or even the current strategy distribution; the observation of aggregate strategy distribution $\bar\x$ is enough for the planner to update $T_i$.

Strictly speaking, in a setting where there are \textit{finitely many} payoff types, \cite{SandholmCongstPr} verified that, with ${\mathbf T}$ being the above dynamic Pigouvian pricing scheme, $\tilde\F^{\mathbf T}$ has a potential function $\tilde f^{\mathbf T}$ being the total payoff: 
$$ \tilde f^{\mathbf T}(\X)=\E_\Theta\left[ \F[\X](\btheta)\cdot\x(\btheta)\right] \quad\text{ with }\X=\int\x d\P_\Theta.$$  
In particular, if the common payoff function $\F^0$ exhibits negative externality, $\tilde f^{\mathbf T}$ is concave and thus the unique social optimum is achieved in the long run through this pricing scheme regardless of the initial state. Thanks to \Cref{thm:NStbl_pot}, now we can extend this claim to the games with \textit{infinitely many} payoff types.\footnote{\citet[p.903]{SandholmNegImpl} speculated it by referring to \cite{ElySandholmBayes1}, which allows us to reduce the heterogeneous standard BRD to a homogeneous smooth BRD of agents by treating persistent heterogeneity in the former as transitory perturbation, i.e., letting an agent draw a new $\btheta$ from $\P_\Theta$ at each revision opportunity. However, \cite{ZusaiDistStbl} finds that  heterogeneous evolutionary dynamics generically cannot reduce to a homogeneous dynamic, except the standard BRD and the smooth BRDs.}
\end{exmp}

\section{Extensions}\label{sec:ext}
\subsection{Observational dynamics}\label{sec:Observ}
In some of major evolutionary dynamics, an agent observes other agents' strategies and the observation influences the agents' switching decision. For example, an agent may imitate other agents' strategies or the switching rate may depend on the relative payoffs compared to the average payoff of the observed population. We can generalize these dynamics as \textit{observational dynamics} by having the strategy distribution among observed agents $\tilde\x\in\Delta^\cS$, not only payoff vector $\bpi\in\R^S$, in the argument of revision protocol $\brho$.

\begin{exmp}
With an \textbf{excess payoff protocol}, a revising agent calculates the average payoff $\tilde\x\cdot\bpi$ and switches to strategy $s'$ with the rate that increases with the excess payoff of the new strategy $\pi_{s'}-\tilde\x\cdot\bpi$. In particular, the revision protocol $\rho_{s s'}(\bpi,\tilde\x)=[\pi_{s'}-\tilde\x\cdot\bpi]_+$ defines the \textbf{Brown-von Neumann-Nash (BNN) dynamic} \citep{Hofbauer00Sel_NashBrownToMaynardSmith}.\footnote{\label{ftnt:ContrStbl_EPT} Excess payoff dynamics allow innovation of a new strategy, while imitative dynamics do not. Thus, stationarity and stability of equilibria are restricted to the interior of the strategy space for the latter dynamics while they are not for the former. Further, it is known that excess payoff dynamics guarantee global asymptotic stability of the Nash equilibrium set in a contractive game (\cite{HofSand09JET_StableGames}; also known as a negative semidefinite game or a stable game), while the latter guarantees only Lyapunov stability unless contractiveness is strict; see \cite{SandholmPopText}. Also, in a continuous strategy space, they result in different characterizations of local stability; see \cite{HofOechsslerRiedel09GEB_BvNDyn_ContStr}.}
\end{exmp}

\begin{exmp}\label{exmp:ImitatDyn}
With an \textbf{imitative protocol}, a revising agent randomly picks another agent and switches to the observed agent's strategy $s'$ with the rate $I_{s s'}(\bpi)\in\R_+$: the overall switching rate is $\rho_{s s'}(\bpi,\tilde\x)=\tilde x_{s'} I_{s s'}(\bpi)$. There are several imitative protocols that yield the \textbf{replicator dynamic} \citep{TaylorJonker78MathBio_ESS}: imitative pairwise comparison $I_{s s'}=[\pi_{s'}-\pi_s]_+$ \citep{Schlag98JET_Imitate}, imitation driven by dissatisfaction $I_{s s'}=\bar\pi-\pi_s$ with constant $\bar\pi\in\R$ \citep{BjonerWeibull96_ImitateEvol}, and imitation of success $I_{s s'}=\pi_{s'}-\underline{\pi}$ with constant $\underline{\pi}\in\R$ \citep{Hof95_ImitDyn}.
\end{exmp}

They fall into L-continuous revision protocols and satisfy \Cref{ass:bddR}.\footnote{Precisely for observational dynamics, $\bar\rho$ is an upper bound on $\rho_{s s'}(\F[\bar\m](\btheta),\m(\btheta))$. As $\rho$ has two arguments, its Lipschitz continuity (under $L^1$ norm on finite-dimensional vectors; see footnote \ref{ftnt:L1norm}) should mean the existence of $L_\rho$ such as $|\rho_{s s'}(\bpi,\tilde\x)-\rho_{s s'}(\bpi',\tilde\x')|\le L_\rho |(\bpi,\tilde\x)-(\bpi',\tilde\x')|=L_\rho (|\bpi-\bpi'| + |\tilde\x-\tilde\x'|)$ for any $s,s'\in\cS, \bpi,\bpi'\in\R^S, \tilde\x,\tilde\x'\in\Delta^\cS$. Technically, we can allow $\rho_{\cdot\cdot}$ to depend on $\X\in\cX$, not only on $\tilde\x\in\Delta^\cS$. Then, this Lipschitz continuity condition is simply generalized as $|\rho_{s s'}(\bpi,\tilde\x)-\rho_{s s'}(\bpi',\tilde\x')|\le L_\rho (|\bpi-\bpi'| + \|\X-\X'\|^\infty_{\cS\Theta})$. We can readily confirm that the proof for \Cref{thm:Lcont_dyn} needs no change, just by glancing over calculation in Supplementary Note \ref{supp:Lcont_dyn}. } (Note that \Cref{ass:cont_type} is not needed for L-continuous revision protocols.) 
~We can readily extend all the positive results, i.e., \Cref{thm:Lcont_dyn,thm:NStat,thm:NStbl_pot}
, to observational dynamics, if we assume that an agent observes the strategy distribution of the same type: a type-$\btheta$ agent observes $\x(\btheta)\in\Delta^\cS$.\footnote{\Cref{cor:NStbl_pot_SpecifDyn} holds for excess payoff dynamics. Imitative dynamics such as the replicator dynamic satisfy the best response stationarity only if $\x^0$ is in the interior of $\Delta^\cS$, and thus these theorems hold for imitative dynamics in the interior of $\cX$.} This assumption of within-type observability matches with an assumption on imitative dynamics in the society of finitely many subpopulations where a member of each subpopulation imitates the behavior of those in the same subpopulation; for example, \cite{Hummel_McAfee_18IER_EvolCons_MonopExit} adopt this assumption.\footnote{However, \cite{Hummel_McAfee_18IER_EvolCons_MonopExit} simply apply a general formula of the replicator dynamic to the evolution of type-conditional strategy distribution and do not construct the dynamic from a revision protocol; thus they do not explicitly discuss imitation.} The proofs of these theorems in the appendix are indeed written explicitly to include $\x(\btheta)$ as an argument of revision protocol $\brho$. 

To maintain the existence of a unique solution trajectory (\Cref{thm:Lcont_dyn}) and stationarity of an equilibrium strategy distribution (\Cref{thm:NStat}), this assumption of within-type observability can be replaced with an alternative assumption that an agent observes the aggregate strategy distribution $\bar\x=\X(\Theta)$, instead of $\x(\btheta)$, and applies the agent's own current payoff vector $\F[\X](\btheta)$ as $\bpi$.\footnote{In an ASAG, it could be done without assuming that an agent precisely knows $\F[\X](\btheta)$, if an agent learn only the common payoffs $\F^0[\X]$ from sampled agents and then distorts it by the agent's own idiosyncratic payoffs $\btheta$.} But, then PC may not be extended from the homogeneous setting to the heterogeneous setting. If observations are sampled from the entire population, stability analysis becomes essentially different from how we have investigated stability in this paper.\footnote{About unobservable heterogeneity in aspiration levels in imitative dynamics, \cite{SawaZusai_HeteroAsp} verify that, although the dynamic becomes more complicated and basic properties such as positive correlation do not hold, long-run outcomes are robust to the introduction of unobservable heterogeneity. For this, they verify that the difference in the aggregate strategy distribution between under the heterogeneous dynamic and under the homogeneous dynamic vanishes in the long run in any game, whether the dynamic converges to equilibrium or not.}

\subsection{Heterogeneity in revision protocols} 
All of our results are robust to heterogeneity in revision protocols. Now, let each type $\btheta\in\Theta$ of agents not only have its peculiar payoff function $\F(\btheta)$ but also follow its own revision protocol $\brho^\btheta$; in the case of an exact optimization protocol, this should be constructed from the conditional switching rate function $(Q^\btheta_{s s'})_{(i,j)\in\cS^2}$. The mean dynamic $\v^\btheta:\R^S\times\Delta^S\to \Delta^S$ is defined by tagging $\btheta$ to \eqref{eq:Dyn_x} as 
$$ \dot x_s(\btheta) =v_s^\btheta(\bpi(\btheta),\x(\btheta)):= \sum_{s'\in\cS} x_{s'}(\btheta) \rho_{s' s}^\btheta(\bpi(\btheta))  - x_s(\btheta)\sum_{s'\in\cS} \rho_{s s'}^\btheta(\bpi(\btheta))  \qquad\text{ for each $s\in\cS$}.$$
Then, heterogeneous dynamic $\v^\F$ is defined in the same fashion as 
$ \dot\x(\btheta)=\v^\F[\x](\btheta):=\v^\btheta(\F[\E_\Theta\x](\btheta),\x(\btheta)).$
Again, these notations are explicitly shown in the proofs in the appendix. \Cref{thm:NStat,thm:NStbl_pot}
~ hold as long as the assumptions in each theorem are satisfied with $\v^\btheta$ of (almost) every type $\btheta\in\Theta$. 

The existence of a unique solution trajectory (\Cref{thm:Lcont_dyn}) is also guaranteed, though we should clarify what the assumptions (including the Lipschitz continuities assumed in \Cref{dfn:ContRev,dfn:ExactOpt}) are imposed on heterogeneous revision protocols. For this, let $\Theta_C$ be the set of types that adopt any of L-continuous revision protocols and $\Theta_E$ be the set of those who use exact optimization protocols with any conditional switching rate functions. Then, the assumptions for \Cref{thm:Lcont_dyn} should read as follows.
\begin{description}
 \item[\Cref{dfn:ContRev}] There should be a \textit{common } Lipschitz constant $\bar L_\rho$ of the switching rate function over almost all the types in $\Theta_C$: $|\rho_{s s'}^\btheta(\bpi)-\rho_{s s'}^\btheta(\bpi')|\le \bar L_\rho |\bpi-\bpi' |$ for any $s,s'\in\cS, \bpi,\bpi'\in\R^S$ and $\P_\Theta$-almost all $\btheta\in\Theta_C$.
 \item[\Cref{dfn:ExactOpt}] There should be a \textit{common } Lipschitz constants $\bar L_Q$ of the conditional switching rate functions $Q_{s s'}^\btheta$ over almost all the types in $\Theta_E$: $|Q_{s s'}^\btheta(\bpi)-Q_{s s'}^\btheta(\bpi')|\le \bar L_Q |\bpi-\bpi' |$ for any $s,s'\in\cS, \bpi,\bpi'\in\R^S$ and $\P_\Theta$-almost all $\btheta\in\Theta_E$.
 \item[\Cref{ass:F0}] Since this is about payoff types, this needs no modification.
 \item[\Cref{ass:bddR}] There should be a \textit{common }upper bound $\bar\rho$ on the switching rate functions $\rho_{s s'}^\btheta$ over almost all the types: $\rho_{s s'}^\btheta(\F[\M](\btheta))\le \bar\rho$ for any truncated finite signed measure $\M\in\bar\cM_{\cS\Theta}$, any $s,s'\in\cS$ and, $\P_\Theta$-almost all $\btheta\in\Theta$.
 \item[\Cref{ass:cont_type}] The assumption is needed as long as $\P_\Theta(\Theta_E)>0$; otherwise, it is not needed.
\end{description}

Of course, this extension to heterogeneous revision protocols cover observational dynamics. While we assume within-type observability for Nash stationarity and stability, it is not needed for the existence of a unique solution trajectory. Thus, our existence theorem would provide the most fundamental starting point to study the effect of both observable and unobservable heterogeneity of revision protocols on population dynamics and equilibrium stability.

\section{Concluding remarks}\label{sec:concl}
In this paper, we extend evolutionary dynamics to allow (possibly) continuously many types under persistent heterogeneity in payoff functions and revision protocols. With a rigorous formulation of a heterogeneous evolutionary dynamic as a differential equation over the space of probability measures, we clarify the regularity conditions on the revision protocol, the game and the type distribution to guarantee the existence of a unique solution path from an arbitrary initial state. We extend equilibrium stationarity in general and equilibrium stability in potential games from the homogeneous setting to the heterogeneous setting. This study establishes the foundation to study evolution in heterogeneous populations and opens up a wide field of applications. 

Our result on extension of equilibrium stability in potential games suggests that any admissible dynamics share global stability of the equilibrium \textit{set} and also asymptotic stability of each local maximizer of the potential function. However, different admissible dynamics may have different basins of attraction and thus they may converge to different locally stable equilibria when starting from the same initial state. Especially, in aggregate games, the preceding studies \citep{ElySandholmBayes1,Blonski99GEB_AnonymousGame_binary} assume \textit{aggregability } in the sense that the dynamic of aggregate state is completely predictable from its current state, independently of the underlying joint strategy distribution over different types. However, \cite{ZusaiDistStbl} argues that evolutionary dynamics are generically nonaggregable, except the standard and smoothed BRDs, even in aggregate games.  On the positive side, it proposes to use nonaggregability to select aggregate equilibria by requiring robustness of stability to any distortion in the underlying strategy distribution under nonaggregable dynamics.\footnote{\cite{ZusaiGains} provides a universal (and economically intuitive) proof for equilibrium stability of a contractive game under a wide range of ``economically reasonable'' dynamics (see the paper for its meaning) by discovering a universal formula of a Lyapunov function; for a contractive game, we need to create a Lyapunov function for each dynamic, as listed in \cite{HofSand09JET_StableGames}. The universal proof suggests that the stability holds robustly under heterogeneous populations, though he restricts attention to a finitely many populations.}


In this paper, we confine our analysis to a game with finitely many strategies while allowing infinitely many types. The formulation and techniques for our study are borrowed from those for evolution in games with continuously many strategies, though it brings new issues as remarked in \Cref{sec:Exist}. While we might be able to use similar techniques based on measure theory, we may need different regularity conditions for evolution on infinite-dimensional type and strategy space. It is left to future research.

%

\section*{Acknowledgment}
I greatly acknowledge Dimitrios Diamantaras, Bill Sandholm, Marek Weretka, and Noah Williams for encouragement and advice, as well as the two anonymous referees and the associate editor of this journal, Man Wah Cheung, Ian Dobson, Shota Fujishima, Hon Ho Kwok, George Lady, Catherine Maclean, Daisuke Oyama, Moritz Ritter, Ryoji Sawa, Marciano Siniscalchi, Douglas Webber and Jiabin Wu for their comments. I also thank Olena Berchuk for careful reading and editing suggestions in an earlier version. This paper was developed from a half of the working paper previously circulated as ``Nonaggregable evolutionary dynamics under payoff heterogeneity:" another half is now separated to \cite{ZusaiDistStbl}. 


\bibliography{../DZbib}

\newpage\appendix
\setcounter{equation}{0}
\renewcommand{\theequation}{\Alph{section}.\arabic{equation}}
\section{Appendix to Section \ref{sec:Model}\label{apdx:model}}
\subsection{Measure-theoretic definition of strategy distribution}\label{sec:comp_Bayes}
This subsection provides mathematically rigorous definitions of a joint strategy distribution and a conditional strategy distribution based on measure theory. Proofs in the appendices deal with joint strategy distributions to properly utilize the Lyapunov stability theorem (\Cref{thm:Cheung_LypnvAsymptStbl} in Appendix \ref{apdx:Exist}) and to borrow measure-theoretic construction of evolutionary dynamics on a continuous strategy space as in \cite{OechsslerRiedelET01_InfStrEvolDyn, OechsslerRiedelJET02_EvolStbl_Cont} and \cite{Cheung13_PairwiseCompDyn_ContStr}. 

Let $\Omega:=[0,1]\subset\R$ be the set (population) of agents. We define a (probability) measure $\P_\Omega:\cB_\Omega\to[0,1]$ as the Lebesgue measure so $\P_\Omega(\Omega)=1$. Denote by $\cB_\Omega$ the Lebesgue $\sigma$-field over $\Omega$.  Let $\s(\omega)\in\cS$ denote the strategy taken by agent $\omega$. We restrict strategy profile $\s:\Omega\to\cS$ to a $\cB_\Omega$-measurable function.

Let $\btheta(\omega)\in\Theta$ be the type of agent $\omega\in\Omega$; recall that type space $\Theta$ is a complete separable space with metric $d_\Theta$. Denote by $\cB_\Theta$ be the Borel $\sigma$-field on this metric space. Agents' type profile $\btheta:\Omega\to\R^T$ is assumed to be measurable with respect to $\cB_\Omega$. Then, it induces probability measure $\P_\Theta:\cB_\Theta\to[0,1]$ by $\P_\Theta(B_\Theta):=\P_\Omega(\{\omega\in\Omega : \btheta(\omega)\in B_\Theta\})$ for each $B_\Theta\in\cB_\Theta$. 

Combination of strategy profile $\s:\Omega\to\cS$ and type profile $\btheta:\Omega\to\Theta$ generates a finite measure $X_s:\cB_\Theta\to\R_+$ for each $s\in\cS$ from $\P_\Omega$:
$$ X_s(B_\Theta):=\P_\Omega(\{\omega\in\Omega ~:~ \s(\omega)=s \text{ and }\btheta(\omega)\in B_\Theta \})\qquad\text{for each $B_\Theta\in\cB_\Theta$}.$$ 
$X_s(B_\Theta)$ represents the mass of strategy-$s$ players whose types belong to set $B_\Theta.$ 
The \textbf{(joint) strategy distribution} $\X$ is collection of these measures $X_s$, i.e., $\X=(X_s)_{s\in\cS}$. We can see this vector measure as a joint probability measure over the product space $\cS\times\Theta$. \footnote{Abusing notation, we could say that $\X$ defines a measure of a Borel set $B_{\cS\Theta}$ on the product space $\cS\times\Theta$ by 
$$ \X(B_{\cS\Theta}):= \sum_{s\in\cS} X_s( \{\btheta\in\Theta : (a,\btheta)\in B_{\cS\Theta}\})=\P_\Omega ( \{\omega\in\Omega : (\s(\omega),\btheta(\omega))\in B_{\cS\Theta} \} ).$$} 
The space of joint strategy distributions $\cX$ is thus the set of probability measures over $\cS\times\Theta$ such that the marginal distribution of types coincides with $\P_\Theta$, i.e., $\sum_{s\in\cS} X_s(B_\Theta)=\P_\Theta(B_\Theta)$ for each $B_\Theta \in \cB_\Theta$. Let $\cB_{\cS\Theta}$ be the Borel $\sigma$-field on the product space $\cS\times\Theta$. 

Since $\X$ must satisfy $X_s(B_\Theta)\le \P_\Theta(B_\Theta)$ for each $s\in\cS$, $X_s$ is dominated by $\P_\Theta$ in the sense that
\begin{equation}
 \P_\Theta(B_\Theta)=0 \quad \Longrightarrow \quad X_s(B_\Theta)=0 \qquad\text{for each $B_\Theta\in\cB_\Theta$}. \label{eq:dfn_sbscont}
\end{equation}
Denote by $X_s\ll\P_\Theta$ this dominance relation, i.e., absolute continuity of $X_s$ with respect to $\P_\Theta$.
It follows by Radon-Nikodym theorem that there exists a $\cB_\Theta$-measurable nonnegative function $x_s:\Theta\to\R_+$ such that 
$$ X_s(B_\Theta)=\int_{B_\Theta} x_s(\btheta) \P_\Theta(d\btheta) \qquad \text{for any $B_\Theta\in\cB_\Theta$}.$$
$x_s$ is the density function of measure $X_s$. The density is determined uniquely in the sense that, if another measurable function $x'_s$ satisfies $X_s(B_\Theta)=\int_{B_\Theta} x'_s(\btheta)\P_\Theta(d\btheta)$ for all $B_\Theta\in\cB_\Theta$, then $x'_s(\btheta)=x_s(\btheta)$ for $\P_\Theta$-almost all $\btheta\in\Theta$.

$\X$ is dominated by $\P_\Theta$ in the sense that $X_s\ll\P_\Theta$ for all $s\in\cS$; we abuse notation to denote this domination by $\X\ll\P_\Theta$. Note that the dominance of joint strategy distribution $\X$ by the type distribution $\P_\Theta$ is peculiar to heterogeneous dynamics, making a difference in the proof of Lipschitz continuity of the dynamic from the one for continuous strategy dynamics. See \Cref{rmk:Lcont_NeedDensity} in \Cref{sec:Exist}.

The collection of Radon-Nikodym densities $\x=(x_s)_{s\in\cS}:\Theta\to\R^S_+$ is the \textit{type-conditional strategy distribution} corresponding to $\X$. From the fact that $\sum_{s\in\cS}X_s(B_\Theta)=\P_\Theta(B_\Theta)$ and $X_s(B_\Theta)\ge 0$ for any $B_\Theta\in\cB_\Theta$ and $s\in\cS$, we can confirm that $\x(\btheta)$ is a probability vector for almost all types:
$$ \x(\btheta)\in\Delta^S \qquad \text{for $\P_\Theta$-almost all $\btheta\in\Theta$.}$$

\subsection{Topology of the space of joint strategy distributions}\label{apdx:norm}
Choice of a topology is a sensitive issue when we argue dynamics of a probability measure over a continuous space. We follow the convention in the literature on evolutionary dynamics over a continuous strategy space, such as in \cite{Cheung13_PairwiseCompDyn_ContStr}. That is, we use the \textit{strong topology} to prove the existence of a unique solution path and the \textit{weak topology} to obtain stability of equilibrium strategy distribution. See \citet[Section 4]{Cheung13_PairwiseCompDyn_ContStr} for a detailed explanation on the strong and weak topology in evolutionary dynamics on a continuous space.

Below we define these two topologies on the space of finite signed measures $\cM_{\cS\Theta}$. Note that $\cX\subset \cM_{\cS\Theta}$ and that $\cM_{\cS\Theta}$ is the tangent space of $\cX$. This space $\cM_{\cS\Theta}$ is a vector space and a transition vector stays in this extended space.

The \textbf{strong topology} is metrized by the variational norm $\| \cdot\|^\infty_{\cS\Theta}$ defined as 
$$  
\| \M \|^\infty_{\cS\Theta}=\sup_{\g}\left\{ \left| \sum_{s\in\cS}  \int_{\btheta\in\Theta} g_s(\btheta) M_s(d\btheta)\right| ~:~ \sup_{(a,\btheta)\in\cS\times\Theta} |g_s(\btheta)|\le 1 \right\},
$$
where the first sup is taken over the set of measurable functions $\g=(g_s)_{s\in\cS}$ on $(\cS\times\Theta, \cB_{\cS\Theta})$. 
Conditional strategy distributions belong to $\cF_\cX$, i.e., the space of $\cB_\Theta$-measurable vector functions from $\Theta$ to $\Delta^S$. Note that, if $\M\ll \P_\Theta$, then there uniquely exists a Radon-Nikodym density $\m\in\cF_\cX$ such that $\M=\int \m d\P_\Theta$ in the sense we defined in Appendix \ref{sec:comp_Bayes}. The theorem below suggests that the variational norm on $\cX$ is equivalent to the $L^1$-norm on $\cF_\cX$.\footnote{\citet{ElySandholmBayes1} define the standard BRD under payoff heterogeneity directly as a dynamic of $\x\in\cF_\cX$ and adopt $L^1$ norm on $\cF_\cX$.} 
The proof is provided in Section  \ref{supp:VarNorm} of Supplementary Note.\footnote{This density-based formula of the variational norm comes essentially from Theorem 5 in \citet{OechsslerRiedelET01_InfStrEvolDyn}.}

\begin{thm}\label{thm:VarNorm}
For any finite signed measure $\M\in\cM_{\cS\Theta}$ with density $\m=(m_s)_{s\in\cS}$, we have
\begin{equation}
\|\M\|^\infty_{\cS\Theta}=\E_{\Theta} |\m(\btheta)|=\int_{\Theta}\sum_{s\in\cS} |m_s(\btheta)| \P_\Theta(d\btheta). \label{Dfn_norm}
\end{equation}
\end{thm}

With the variational norm, the normed vector space $(\cM_{\cS\Theta}, {\|\cdot\|^\infty_{\cS\Theta}})$ is a Banach space; but not with weak topology. By \citet[Cor. 3.9]{Zeidler86}, boundedness and Lipschitz continuity of the dynamic in the strong topology jointly guarantee the existence and uniqueness of a solution path of the dynamic. See \Cref{thm:Zeidler_SolExist} in \Cref{apdx:Lcont_dyn}.

Under the \textbf{weak topology} on the set of measures over space $\cS\times\Theta$, a mapping from $\cM_{\cS\Theta}\to\R$ such as $\mu\mapsto \int_\cS f d\mu$ is continuous for any bounded and continuous function $f:\cS\times\Theta\to\R$. The product space $\cS\times\Theta$ is separable with metric $d_{\cS\Theta}:(\cS\times\Theta)^2\to\R_+$ given by\footnote{%
The metric $d_{\cS\Theta}$ is a product metric constructed from the discrete norm on $\cS$ and metric $d_\Theta$ on $\Theta$. Notice $S<\infty$ and $\Theta$ is separable under $d_\Theta$; so the product metric $d_{\cS\Theta}$ makes $\cS\times\Theta$ separable. Here $\bone\{s\ne s'\}$ is an indicator function and takes 1 if $s\ne s'$ and 0 otherwise.}
$$ d_{\cS\Theta}((s,\btheta),(s',\btheta')):=\bone\{s\ne s'\}+ d_\Theta(\btheta,\btheta').$$ 
Then, the weak topology is metrized by Prokhorov metric $d_\cM:{\cM_{\cS\Theta}}^2\to\R_+$ such that\footnote{%
If there is no payoff heterogeneity, i.e., $\Theta=\{\btheta_0\},$ then strategy distribution $\M$ can be simply represented by an $A$-dimensional vector $(\bar m_s)_{s\in\cS}\in\R^S$ such that $\bar m_s=M_s(\{\btheta_0\})$. Then, $d_\cM(\M,\M')=\varepsilon$ is equivalent to $\sup_{s\in\cS} |\bar m_s -\bar m'_s| =\varepsilon.$ So the metric $d_\cM$ reduces to the sup norm on $\R^S$.}
\begin{align*}
d_\cM (\M,\M'):=\inf\{ \varepsilon >0 ~:~ &\M(B_{\cS\Theta})\le \M'(B_{\cS\Theta}^\varepsilon) +\varepsilon \\
	& \text{ and }\M'(B_{\cS\Theta})\le \M(B_{\cS\Theta}^\varepsilon) +\varepsilon \quad \text{ for all } B_{\cS\Theta}\in\cB_{\cS\Theta} \},
\end{align*}
where $B_{\cS\Theta}^\varepsilon$ is defined from $B_{\cS\Theta}$ as $B_{\cS\Theta}^\varepsilon:=\{ (s,\btheta)\in \cS\times\Theta ~:~  d_{\cS\Theta}( (s,\btheta), (s',\btheta'))< \varepsilon \text{ with some }  (s',\btheta')\in B_{\cS\Theta} \}$.\footnote{If $\varepsilon<1$, the condition for $(a,\btheta)\in B_{\cS\Theta}^\varepsilon$ is equivalent to the existence of $\btheta'\in\Theta$ such that $d_\Theta(\btheta,\btheta') <\varepsilon$ and $(a,\btheta')\in B_{\cS\Theta}.$} 
Under the weak topology, the space of probability measures, i.e., the space of joint strategy distributions becomes compact. Then, we can apply the Lyapunov stability theorem, as in \citet[Thm. 6]{Cheung13_PairwiseCompDyn_ContStr}. See \Cref{thm:Cheung_LypnvAsymptStbl} in \Cref{apdx:NStbl_pot}.

\section{Appendix to Section \ref{sec:Exist}}\label{apdx:Exist}

\subsection{Sketch of Proof of \Cref{thm:Lcont_dyn}}\label{apdx:Lcont_dyn}
We prove the existence of a unique solution trajectory under a heterogeneous dynamic by verifying it for the corresponding dynamic of the joint strategy distribution, appealing to the equivalence between conditional strategy distributions and joint strategy distributions. Here we sketch the outline of the proof, while the complete presentation of the proof is provided in Section  \ref{supp:Lcont_dyn} of Supplementary Note.

First, we construct the mean dynamic of joint strategy distribution $\V=(V_s)_{s\in\cS}:\cX\times\cC_\Theta\to\cM_{\cS\Theta}$ by gathering the mean dynamic $\v^\btheta$ of a conditional strategy distribution, defined by \eqref{eq:Dyn_x}, over all $\btheta\in\Theta$: for each strategy $s\in\cS$, 
\begin{align}
\dot X_s(B_\Theta) &= V_s[\X,\bpi](B_\Theta) = \int_{B_\Theta} v_s^\btheta[\bpi(\btheta),\x(\btheta)] \P_\Theta(d\btheta) \notag\\
	&=\int_{B_\Theta}\sum_{s'\in\cS} \rho_{s' s}^\btheta(\bpi(\btheta), \x(\btheta)) X_{s'}(d\btheta) - \int_{B_\Theta} \left\{ \sum_{s'\in\cS} \rho_{s s'}^\btheta(\bpi(\btheta), \x(\btheta))\right\} X_s(d\btheta) \label{eq:Dyn_composite}
\end{align}
for each $B_\Theta\in\cB_\Theta$, given joint strategy distribution $\X=\int\x d\P_\Theta\in\cX$ and payoff profile $\bpi:\Theta\to\R^S$. In short, we write $\dot\X=\V[\X,\bpi]$. 

In a population game $\F:\Delta^\cS\times\Theta\to\R^S$, the mean dynamic \eqref{eq:Dyn_composite} of a joint strategy distribution defines an autonomous dynamic $\V^\F$ over $\cX$ by 
$$\dot \X=\V^\F[\X]:=\V[\X,\F[\X]]\in\cM_{\cS\Theta}$$
for each joint strategy distribution $\X\in\cX$. Then, this dynamic of joint strategy distribution $\V^\F$ matches with the dynamic of conditional strategy distribution $\v^\F$ defined in \Cref{sec:dyn},  in the sense that $\V^\F[\X](B_\Theta)=\int_{B_\Theta} \v^\F[\x](\btheta) \P_\Theta(d\btheta),$ where $\x$ is the corresponding conditional strategy distribution, i.e., the Radon-Nikodym density of $\X$. 

To argue the existence of a unique solution trajectory, we exploit the known result on a Lipschitz continuous dynamic over a Banach space as in the theorem below.\footnote{See \citet[Theorem A.3.]{ElySandholmBayes1} for a version of this theorem for $\cF_\cX$, which guarantees the existence of a unique solution trajectory for a heterogeneous dynamic on $\cF_\cX$ with $L^1$-norm from Lipschitz continuity of the dynamic.}
\begin{thm}[\citealp{Zeidler86}: Corollary 3.9]\label{thm:Zeidler_SolExist}
Consider a dynamic $\dot z=V(z)$ with $V:\mathcal Z\to \mathcal Z$. If the space $\mathcal Z$ is a Banach space and the dynamic $V$ is Lipschitz continuous and bounded, then there exists a unique solution $\{z_t\}_{t\in\R_+}$ from any initial state in $z_0\in\mathcal Z$.
\end{thm}

For this, we need a Banach space. But, the space of joint strategy distributions $\cX$ is not a vector space. Thus, we extend the domain of the dynamic to the space of finite signed measures $\cM_{\cS\times \Theta}$. Since the mean dynamic $\V[\X,\bpi](B_\Theta)$ is defined by collecting the transition of  the density $\x(\btheta)$ over types $\btheta\in B_\Theta$, we still need a density of a measure on this extended space. However, a finite signed measure may not be absolutely continuous with respect to the type distribution $\P_\Theta$. We use the Lebesgue decomposition theorem to extract the absolutely continuous part. 

\begin{lem}[\citealp{Rudin_RealComplexAnalysis}: \S 6.10]\label{lem:LebesgueDecomp}
For any finite signed measure $\M=(M_s)_{s\in\cS}\in\cM_{\cS\Theta}$, there is a pair of finite signed measures $\tilde\M=(\tilde M_s)_{s\in\cS},\hat\M=(\hat M_s)_{s\in\cS}\in\cM_{\cS\Theta}$ such that, for each $s\in\cS$,
\begin{enumerate}[i)]
 \item $M_s=\tilde M_s+\hat M_s$;
 \item $\tilde M_s\ll \P_\Theta$, i.e., $\P_\Theta(B_\Theta)=0 \quad \Longrightarrow \quad \tilde M_s(B_\Theta)=0$ for any $B_\Theta\in\cB_\Theta$
 \item $\hat M_s\perp \P_\Theta$, i.e., there exists $E_s\in\cB_\Theta$ such that $\hat M_s(B_\Theta\cap E_s)=0$ and $\P_\Theta(B_\Theta\setminus E_s)=0$ for any $B_\Theta\in\cB_\Theta$.
\end{enumerate}
\end{lem}
The part (ii) implies that $\tilde\M$ has density $\tilde \m=(\tilde m_s)_{s\in\cS}$ with respect to $\P_\Theta$. Besides, $\|\tilde \M\|\le \|\M\|$, since i) and ii) imply $\|\M\|=\|\tilde\M\|+\|\hat\M\|.$
We extend $\V$ to $\cM_{A\times\Theta}$ by discarding the orthogonal part $\hat\M$ and applying \eqref{eq:Dyn_composite} to the continuous part $\tilde\M$. This need for the Lebesgue decomposition is the first difference from evolutionary dynamics on a continuous \textit{strategy }space: see Remark \ref{rmk:Lcont_NeedDensity} in Section \ref{sec:Exist}. Let $\tilde \cM_{\cS\Theta}$ be the space of $\P_\Theta$-absolutely continuous measures. 

Yet, the density function $\tilde\m$ of $\tilde\M$ may not be bounded, while that of a joint strategy distribution, i.e., a conditional strategy distribution $\x$ is bounded in the sense that $\x(\btheta)$ of almost every type $\btheta$ belongs to a bounded set $\Delta^\cS$. We will utilize the assumptions that the payoff function $\F$ and the switching rate function $\brho^\btheta$ (or the conditional switching rate function $Q_{s s'}$ for exact optimization protocols) are continuous and thus bounded if its domain is restricted to a compact set. To restrict the value of the density function into a compact domain, we truncate $\tilde\m(\btheta)$ of each $\btheta$ by a rounding function $\bmu=(\mu_{s'})_{s'\in\cS}:\R^S\to[-3,3]^S$ such that $\bmu(\z)=\z$ if $\z\in\Delta^\cS$ and $\bmu$ is Lipschitz continuous with constant $L_\bmu$.\footnote{For example, define $\mu^0:\R\to [-3,3]$ such as $\mu^0(z):=-3+\exp(z+2)$ for $z<-2$, $\mu^0(z):=z$ for $z\in[-2,2]$ and $\mu^0(z):=3-\exp(2-z)$ for $z>2.$ Then, define vector function $\bmu=(\mu_s)_{s\in\cS}$ by $\mu_s(\z)=\mu^0(z_s)$ for each $s\in\cS$ and $\z\in\Delta^S$.} 

Then, we redefine function $\v^\F:\tilde\cM_{\cS\Theta}\times\Theta\to\R^S$ on the extended domain by 
$$ \v^\F[\tilde\M](\btheta)= \v^\btheta(\F[\E_\Theta\bmu(\tilde\m(\btheta))](\btheta),\bmu(\tilde\m(\btheta)) )$$
for each $s\in\cS$ and any $\P_\Theta$-absolutely continuous finite signed vector measure $\tilde\M\in\tilde\cM_{\cS\Theta}$ with the Radon-Nikodym density $\tilde\m$. This leads to the extension of $\V^\F:=(V^\F_s)_{s\in\cS}$ to $\cM_{\cS\Theta}$, such as 
$$ V_s^\F[\M](B_\Theta)=\int_{B_\Theta} v_s^\F[\tilde\M](\btheta) \P_{\Theta}(d\btheta)$$
for each $s\in\cS$, any $B_\Theta\in\cB_\Theta$ and any finite signed vector measure $\M\in\cM_{\cS\Theta}$; here $\tilde\M$ is the $\P_\Theta$-absolutely continuous part of $\M$ in the Lebesgue decomposition of $\M$. As only this part matters to the value of $\V^\F$, we have $\V^\F[\M]=\V^\F[\tilde\M]$. 

To prove Lipschitz continuity of $\V^\F$, we look at $\V^\F$ on $\tilde\cM_{\cS\Theta}$: in Supplementary Note \ref{supp:Lcont_dyn}, we find $L_V^\F>0$ such that\footnote{Here the norm $\|\cdot\|$ is the variational norm, defined in Appendix \ref{apdx:norm}.}
\begin{equation}
\| \V^\F[\tilde\M]-\V^\F[\tilde\M'] \| \le L_V^\F \| \tilde\M-\tilde\M' \| \qquad \text{ for any }\tilde\M,\tilde\M'\in\tilde\cM_{\cS\Theta}.  \label{eq:Lcont_V}
\end{equation}
Then, this implies Lipschitz continuity over the whole space $\cM_{\cS\Theta}$, because $\|\tilde\M-\tilde\M'\|\le \|\M-\M'\|$ for any $\M=\tilde\M+\hat\M,\M'=\tilde\M'+\hat\M'\in\cM_{\cS\Theta}$.

For an L-continuous revision protocol, the Lipschitz continuity of $\V^\F$ is a natural consequence of the Lipschitz continuity of switching rate function $\brho^\btheta$ and of payoff function $\F$.

On the other hand, an exact optimization protocol is discontinuous. If the best response strategies for some type of agents have changed by a change in the joint strategy distribution from $\tilde\M$ to $\tilde\M'$, these agents should experience discontinuous changes in the switching rates. However, these discontinuous changes in their switching rates are bounded thanks to the boundedness of switching rate function $\brho^\btheta$. Further, thanks to Assumption \ref{ass:cont_type}, the mass of agents who belong to such types increases only (Lipschitz) continuously with the change in the joint strategy distribution.\footnote{Note that this assumption also restricts the mass of types who have multiple best responses to a null set (zero measure) in $\P_\Theta$.} As a result, the aggregate change in their switching rates grows only continuously. This mitigation of discontinuity in an exact optimization protocol by continuity of the type distribution marks the second difference from the preceding studies on continuous-\textit{strategy} evolutionary dynamics: see Remark \ref{rmk:Lcont_DiscontExactOpt} in Section \ref{sec:Exist}.

\section{Appendix to Section \ref{sec:EqmComp_Ext}}\label{apdx:EqmComp_Ext}

\subsection{Proof of \Cref{thm:NStat}}\label{apdx:NStat}
\begin{proof}
First of all, joint strategy distribution $\X=\int \x d\P_\Theta$ being an equilibrium \eqref{EqmComp} is equivalent to the corresponding conditional strategy distribution $\x$ being an equilibrium \eqref{BayesEqm}, i.e., $\x(\btheta)\in \Delta \cS^\F_\BR [\X](\btheta)$ for $\P_\Theta$-almost all types $\btheta$. Then, for such $\btheta$, $\x(\btheta)\in \Delta \cS^\F_\BR [\X](\btheta)$ is equivalent to $\v^\F[\x](\btheta)=\0$ by \eqref{eq:BRS_type}. It holds for $\P_\Theta$-almost all types $\btheta$, which means the stationarity of conditional strategy distribution $\x$. This is equivalent to stationarity of joint strategy distribution $\X$, i.e., $\V^\F[\X]=\O$.
\end{proof}

\subsection{Proof of \Cref{thm:ExistEqm}}\label{apdx:ExistEqm}
First of all, notice that an equilibrium strategy distribution is a fixed point of the ``distributional strategy'' best response correspondence $B:\cX\rightrightarrows \cX$ defined as
$$ B[\X] := \argmax_{\Y\in\cX} \E_\Theta \F[\X]\cdot\Y \qquad\text{ for each }\X\in\cX.$$
Below we prove that the assumptions in \Cref{thm:ExistEqm} assures that the maximized  function 
$$ \E_\Theta \F[\X]\cdot\Y =\int_{\Theta} \F[\X](\btheta)\cdot\Y(d\btheta) $$
is continuous in $(\X,\Y)\in\cX^2$.

\begin{proof}[Proof of continuity of {$\E_\Theta \F[\X]\cdot\Y$}]
Fix $\varepsilon>0$ and $(\X,\Y)\in\cX^2$ arbitrarily. By equicontinuity of $\F$, we have some $\delta_\text{Ct}[\X]>0$ such that, whenever $d_\cM(\X',\X)<\delta_\text{Ct}[\X]$, we have $|F_s[\X'](\btheta)-F_s[\X](\btheta)|<0.5\varepsilon$ for any $s\in\cS$ and $\P_\Theta$-almost all $\btheta$. The latter statement implies, for any $\Y'\in\cX$,
\begin{align*}
& \left| \int_\Theta \left (\F[\X'](\btheta)-\F[\X](\btheta) \right) \cdot\Y'(d\btheta) \right| 
	\le \int_\Theta \sum_{s\in\cS} \left|F_s[\X'](\btheta)-F_s[\X](\btheta) \right| Y'_s(d\btheta) \\
<&\int_\Theta \sum_{s\in\cS} 0.5\varepsilon Y'_s(d\btheta)
= 0.5\varepsilon \int_\Theta \P(d\btheta)= 0.5\varepsilon.
\end{align*}

By near-boundedness of $\F$, there exists a combination of $\bar F [\X]\in\R$ and $\delta^1_\text{Bd}[\X]>0$ such that 
$$  d_\cM(\Y,\Y')<\delta^1_\text{Bd}[\X] \quad\Longrightarrow\quad \left| \int_\Theta \sum_{s\in\cS} \left[ |F_s[\X](\btheta)| -\bar F [\X] \right]_+ \cdot (\Y(d\btheta)-\Y'(d\btheta)) \right| < 0.25\varepsilon.$$
For each $s\in\cS$, define set $\Theta^\varepsilon_s\subset\Theta$ and function $g^\varepsilon_s[\X]:\Theta\to\R_+$ by 
$$ \Theta^\varepsilon_s  := \{\btheta : |F_s[\X](\btheta)| > \bar F [\X]\}, \qquad
g^\varepsilon_s[\X](\btheta) :=\bone\{\btheta\in\Theta^\varepsilon_s\} \bar F [\X]+\bone\{\btheta\notin\Theta^\varepsilon_s\} | F_s[\X](\btheta)|.
$$
Notice that $g^\varepsilon_s[\X](\btheta)+\left[ |F_s[\X](\btheta)| -\bar F [\X] \right]_+ \equiv |F_s[\X](\btheta)|$ for any $\btheta\in\Theta$. Since $F_s[\X]:\Theta\to\R$ is measurable, $\Theta^\varepsilon_s$ is a measurable set and $g^\varepsilon_s[\X]$ is a measurable function on $\Theta$. This function is bounded by definition and also continuous; so is the vector-valued function $\g^\varepsilon[\X]=(g^\varepsilon_s[\X])_{s\in\cS}:\Theta\to\R^S_+.$ Hence, under the weak topology, $\int_\Theta \g^\varepsilon[\X](\btheta)\cdot \M(d\btheta)$ is a continuous function of finite signed measure $\M\in\cM_{\cS\Theta}$; there exists $\delta^2_\text{Bd}[\X]>0$ such that, whenever $\Y'$ satisfies, $d_\cM(\Y',\Y)<\delta^2_\text{Bd}[\X]$, we have satisfies
$$ 0.25\varepsilon > \left| \int_\Theta \g^\varepsilon[\X](\btheta) \cdot \left(\Y'(d\btheta)-\Y(d\btheta)\right) \right| = \left| \int_\Theta \sum_{s\in\cS} g^\varepsilon_s[\X](\btheta) \left(Y'_s(d\btheta)- Y_s(d\btheta)\right) \right|. $$

Therefore, any $\Y'$ such as $d_\cM(\Y',\Y)<\delta_\Y:=\min\{\delta^1_\text{Bd}[\X],\delta^2_\text{Bd}[\X]\}$ satisfies
\begin{align*}
 & \left| \int_\Theta \F[\X](\btheta) \cdot \left(\Y'(d\btheta)-\Y(d\btheta)\right) \right| = \left| \int_\Theta \sum_{s\in\cS} F_s[\X](\btheta) \left(Y'_s(d\btheta)- Y_s(d\btheta)\right) \right|\\
\le &  \left| \int_\Theta \sum_{s\in\cS} |F_s[\X](\btheta)| \left(Y'_s(d\btheta)- Y_s(d\btheta)\right) \right|=\left| \int_\Theta \sum_{s\in\cS} \left( g^\varepsilon_s[\X](\btheta)+\left[ |F_s[\X](\btheta)| -\bar F [\X] \right]_+ \right) \left(Y'_s(d\btheta)- Y_s(d\btheta)\right) \right|\\
\le & \left| \int_\Theta \sum_{s\in\cS} g^\varepsilon_s[\X](\btheta) \left(Y'_s(d\btheta)- Y_s(d\btheta)\right) \right| + \left| \int_\Theta \sum_{s\in\cS} \left[ |F_s[\X](\btheta)| -\bar F [\X] \right]_+ \cdot (\Y(d\btheta)-\Y'(d\btheta)) \right| \\
\le & 0.25 \varepsilon+0.25 \varepsilon=0.5 \varepsilon.
\end{align*}

Therefore, any pair of $(\X',\Y)\in\cX^2$ such that $d_\cM(\X',\X)<\delta_\text{Ct}[\X]$ and $d_\cM(\Y',\Y)<\delta_\Y$ satisfies
\begin{align*}
& \left| \E_\Theta \F[\X']\cdot\Y' -\E_\Theta \F[\X]\cdot\Y \right|
= \left| \int_\Theta \F[\X'](\btheta) \cdot \Y'(d\btheta)-\int_\Theta \F[\X](\btheta) \cdot \Y(d\btheta) \right| \\
\le & \left| \int_\Theta \left (\F[\X'](\btheta)-\F[\X](\btheta) \right) \cdot\Y'(d\btheta) \right| +\left| \int_\Theta \F[\X](\btheta) \cdot \left(\Y'(d\btheta)-\Y(d\btheta)\right) \right|\\
< &0.5 \varepsilon+0.5 \varepsilon=\varepsilon.
\end{align*} 
That is, $\E_\Theta \F[\X]\cdot\Y$ is continuous (with respect to the variational norm) in $\cX^2$ at each $(\X,\Y)\in\cX^2$.
\end{proof} 

\begin{proof}[Proof of \Cref{thm:ExistEqm}] To show the existence of a fixed point of the ``mixed strategy'' best response correspondence $B:\cX\rightrightarrows \cX$, we use Glicksberg's fixed point theorem \cite[henceforth AP; Corollary 17.55]{AliprantisBorder_InfiniteDimAnal}. First, we confirm the assumptions on domain $\cX$. The type space $\Theta$ is a complete separable metric space. Since the domain $\cX$ is regarded as the set of distributional strategies over the product of this type space $\Theta$ and the finite strategy space $\cS$, we can borrow the result in \cite{MilgromWeber85MathOR_DistrStr} about $\cX$:\footnote{We could interpret $\E_\Theta\F[\X]\cdot\Y$ as the ``payoff'' from distributional strategy $\X$ in their finite-player model. However, this is different from their payoff function, which is constructed from a normal-form game and thus is bilinear in $\X$ and $\Y$.} $\cX$ is a nonempty, compact and convex subspace of $\cM_{\cS\Theta}$, which is convex and Hausdorff under the weak topology.

With nonemptiness and compactness of $\cX$, continuity of the maximized function $\E_\Theta \F[\X]\cdot\Y$ implies by Berge's maximum theorem (AP, Theorem 17.31) that $B$ is nonempty, compact-valued and upper hemicontinuous. In the Hausdorff metric space, this further implies by AP Theorem 17.10 that $B$ has a closed graph. Since $\E_\Theta \F[\X]\cdot\Y$ is a linear function of $\Y$, $B$ is convex-valued. From the aforementioned properties of $\cX$ and these properties of $B$, Glicksberg's fixed point theorem guarantees the existence of a fixed point of $B$ (as well as compactness of the set of fixed points). 
\end{proof}
\subsection{Proof of \Cref{thm:potExmp}}\label{apdx:potExmp}
\begin{proof} 
Here we prove the claim for \Cref{exmp:ASAG_Dfn}$'$. Proofs for the other examples can be done in a similar fashion and thus they are relegated to Supplementary Note. As $f(\X)=f^0(\X(\Theta))+\E_\Theta[\btheta\cdot \x(\btheta)]$, weak continuity of $f$ is obtained from continuity of $f^0$ and the dominated convergence theorem. 
By applying the definition of $f$ to $\X+\Delta\X$, we have
\begin{align*}
f(\X+\Delta\X) &= f^0(\bar\x+\Delta\bar\x) +\int_\Theta \sum_{s\in\cS} \theta_s (X_s +\Delta X_s)(d\btheta) \\
	&= \left\{ f^0(\bar\x)+ \nabla f^0(\bar\x)\cdot \Delta\bar\x +o(|\Delta\bar\x|)\right\} +\left\{ \int_\Theta \sum_{s\in\cS} \theta_s X_s (d\btheta) + \int_\Theta \sum_{s\in\cS} \theta_s \Delta X_s(d\btheta) \right\}\\
	&= f(\X)+\F^0(\bar\x)\cdot\Delta\bar\x+ \int_\Theta \sum_{s\in\cS} \theta_s \Delta X_s(d\btheta)+o(|\Delta\bar\x|)\\
	&= f(\X)+ \int_\Theta \sum_{s\in\cS} (F^0_s(\bar\x)+\theta_s)  \Delta X_s(d\btheta)+o(|\Delta\bar\x|).
\end{align*}
Here $\bar\x=\X(\Theta)$ and $\Delta\bar\x=\Delta\X(\Theta)$. The second equality comes from differentiability of $f^0$; the third is from the assumption that $f^0$ is a potential function of $\F^0$ and the definition of $f$ applied to $\X$. Then, we should recall $F_s[\X(\Theta)](\btheta)=F^0_s(\X(\Theta))+\theta_s$. So the second term is $\langle \F[\X(\Theta)], \Delta\X \rangle$. About the third error term, note that $|\Delta\bar\x|=|\Delta\X(\Theta)|\le \|\Delta\X\|$. Therefore, we obtain 
$$ f(\X+\Delta\X)=f(\X)+\langle \F[\X(\Theta)], \Delta\X \rangle+o(\|\Delta\X\|).$$
Thus, $f$ is (Fr\'{e}chet) differentiable with derivative $\nabla f(\X)\equiv \F[\X(\Theta)]$. So we have verified that $f$ is a potential function of the game $\F$ defined on $\cX$.
\end{proof}

\subsection{Proof of \Cref{thm:NStbl_pot}}\label{apdx:NStbl_pot}
For stability, we use the weak topology and apply the Lyapunov stability theorem, as in \citet{Cheung13_PairwiseCompDyn_ContStr}. 
\begin{thm}[\citealp{Cheung13_PairwiseCompDyn_ContStr}: Theorems 5--6, Corollary 2]\label{thm:Cheung_LypnvAsymptStbl}
Let $Z\subset \cX$ be a closed set and let $Y\subset \cX$ be a neighborhood of $Z$ in the weak topology on $\cX$. Let $\mathcal L:Y\to\R$ be a decreasing Lyapunov function for dynamic $\V$: that is, $\mathcal L$ is continuous with respect to the weak topology and Fr\'{e}chet-differentiable with $\dot{\mathcal L}(\X)=\langle \nabla \mathcal L(\X), \V[\X]\rangle\le 0$ for all $\X\in Y$. Then, the following holds.
\begin{enumerate}[i)]
 \item Any solution path starting from $Y$ converges to the set $\{\X\in Y \mid \dot{\mathcal L}(\X)=0 \}$ with respect to the weak topology; i.e., this set is attracting under $\V$.
 \item If $\mathcal L^{-1}(0)=Z$, $Z$ is Lyapunov stable under $\V$ with respect to the weak topology. Furthermore, if $\dot{\mathcal L}(\X)<0$ whenever $\X\in Y\setminus Z$, then $Z$ is asymptotically stable under $\V$. 
\end{enumerate}
\end{thm}
Part i) holds for an increasing Lyapunov function; part ii) is retained by defining $Z$ as an isolated set of local maxima.

\begin{proof} \textbf{i)} Since $f$ is a potential function for $\F$, we have
$$ \dot f(\X)=\langle \nabla f(\X), \dot\X \rangle=\langle \F[\X], \V^\F[\X]\rangle =\E_\Theta \left[ \F[\X](\btheta)\cdot\v^\F[\x](\btheta) \right],$$
where $\bar\x=\X(\Theta)$ and $\X=\int_\Theta \x d\P_\Theta$.

Since $\v^\F[\x](\btheta)=\v^\btheta(\F[\X](\btheta),\x(\btheta))$, the first part of \eqref{eq:PC_type} implies $\F[\X](\btheta)\cdot\v^\F[\x](\btheta)\ge 0$ for all $\btheta$ and thus 
$$ \dot f(\X)=\E_\Theta \left[ \F[\X](\btheta)\cdot\v^\F[\x](\btheta) \right]\ge 0.$$

Suppose $\X$ is not a stationarity state under dynamic $\V^\F$, which is equivalent to $\v^\F[\x](\btheta)\ne\0$ for $\P_\Theta$-almost all types $\btheta$ . For a type with $\v^\F[\x](\btheta)\ne\0$, the second part of  \eqref{eq:PC_type} implies $\F[\bar\x](\btheta)\cdot\v^\F[\x](\btheta)>0$. Since this holds for a positive mass of types, we have 
$$ \dot f(\X)=\E_\Theta \left[ \F[\bar\x](\btheta)\cdot\v^\F[\x](\btheta) \right]>0.$$
Therefore, $f$ is a strictly increasing Lyapunov function and the set $\{\X\in\cX : \dot f(\X)=0 \}$ is the set of stationary states, i.e.,$\{\X\in\cX : \V^\F[\X]=\O \}$. By \Cref{thm:Cheung_LypnvAsymptStbl}, this implies that the set of stationary states is globally attracting; a local maximum (local strict maximum, resp.) of $f$ is Lyapunov stable (asymptotically stable, resp.).

\paragraph{ii)} a) Suppose that the corresponding isolated stationary strategy distribution $\X^\ast$ is asymptotically stable, with a basin of attraction $\cX^0\subset\cX^\ast$. Take an arbitrary joint strategy distribution $\X_0$ from $\cX^0$ and let $\{\X_t\}_{t\in\R_+}$ be a solution trajectory under the heterogeneous dynamic from $\X_0.$ Since $f$ is a strictly increasing Lyapunov function, it must be the case that $\dot f(\X_t)>0$ as long as $\X_t$ has not reached exactly $\X^\ast$. Thus, $f(\X^\ast)=f(\X_0)+\int_0^\infty \dot f(\X_t)dt>f(\X_0)$. Since $\X_0$ is taken arbitrarily from $\cX^0$, this verifies that $\X^\ast$ strictly maximizes $f$ in this neighborhood $\cX^0$.

b) We prove the claim by contradiction. Assume that, while $\X^\ast$ is Lyapunov stable, $\X^\ast$ is not a local maximum of $f$. Take an open neighborhood $\tilde\cX^\ast$ of $\X^\ast$ such that $\cl\tilde\cX^\ast\subset \cX^\ast$. Let $\cX^1:=f^{-1}((f(\X^\ast)-h,f(\X^\ast)+h))\cap \tilde\cX^\ast$ with an arbitrarily fixed $h>0$; since $f$ is continuous, $\cX^1$ is an open neighborhood of $\X^\ast$. By Lyapunov stability of $\X^\ast$, there exists another open neighborhood $\cX^0\subset \cX^1$ of $\X^\ast$ such that any solution trajectory starting from $\cX^0$ stays in $\cX^1$ at any moment of time. As $\X^\ast$ is not a local maximum, there exists another strategy distribution $\X^\dagger\in \cX^0$ such that $f(\X^\dagger)>f(\X^\ast)$; note that $f(\X^\dagger)<f(\X^\ast)+h$ since $\X^\dagger\in\cX^0\subset\cX^1$.

Let $\bar\cX^2=f^{-1}([f(\X^\ast)-h,f(\X^\ast)+h])\cap \cl \tilde\cX^\ast$; $\cX^1\subset \bar\cX^2$ and this is closed in $\cX$ and thus compact, as $\cX$ is compact. Consider a solution trajectory $\{\X^t\}_{t\in\R_+}$ starting from $\X^\dagger$; it stays in $\cX^1$ and thus in $\bar\cX^2$. By PC, $f(\X^t)$ is weakly increasing in $t$ and thus $f(\X^t)\ge f(\X^0)=f(\X^\dagger)>f(\X^\ast)$. As $\X^\ast$ is the only stationary point in $\cX^\ast\supset \bar\cX^2$, this implies that the trajectory $\{\X^t\}$ does not arrive at or even converge to a stationary point; thus, PC further implies $f(\X^t)$ \textit{strictly }increases in $t$. Since $f(\X^t)\in [f(\X^\dagger), f(\X^\ast)+h]]$ (the upper bound obtained by $\X^t\in\bar\cX^2$) for all $t$, $f(\X^t)$ must converge to some finite value $\bar f\in [f(\X^\dagger), f(\X^\ast)+h]$. The value of the potential $f(\X^t)$ is expressed as 
$$ f(\X^t)=\int_0^t \dot f(\X_t) dt=\int_0^t \gamma(\X_t) dt$$
with $\gamma(\X)=\langle \F[\X],\V^\F[\X] \rangle$. 

Define $\tilde\gamma:\R_+\to\R$ by $\tilde\gamma(t)=\gamma(\X^t)$. As $\gamma(\X)$ is continuous in $\X$ and $\X^t$ is continuous in $t$, $\gamma(\X^t)$ is continuous in $t$. With the convergence of $f(\X^t)=\int_0^t \tilde\gamma(\tau)d\tau$, this implies there exists a sequence $\{t_n\}_{n\in\N}$ such that $t_n\to\infty$ and $\tilde\gamma(t_n)\to 0$.\footnote{Consider continuous function $g:\R_+\to\R_+$ that always takes a non-negative value. Improper integral $\int_0^\infty g(\tau)d\tau$ is well-defined (i.e., converges to a finite real number) if and only if, for any $\varepsilon>0$, there exists $T\ge 0$ such that $|\int_t^{t'} g(\tau)d\tau |<\varepsilon $ for any $t'>t>T$; fix $T$ to the one that satisfies this for $\varepsilon=1$. Define a sequence $\{t_n\}_{n\in\N}$ by letting $t_n$ be the moment of time to attain the minimum value of $g$ in close interval of time $[T+2^n+1, T+2^{n+1}]$, i.e., $g(t_n)=\min_{\tau\in [T+2^n+1, T+2^{n+1}]} g(\tau)$; this minimum exists and $t_n\to\infty$, since $g$ is continuous and we have $t_n\le T+2^{n+1}\le T+2^{n+1}+1\le t_{n+1}$. For each $n\in\N$, this $t_n$ satisfies $0\le (2^{n+1}-2^n-1)g(t_n)\le \int_{T+2^n+1}^{T+2^{n+1}} g(\tau)d\tau=|\int_{T+2^n+1}^{T+2^{n+1}} g(\tau)d\tau |<1$; hence, $0\le g(t_n)\le 1/ (2^{n+1}-2^n-1)=1/(2^n-1)$. This implies $g(t_n)\to 0$ as $n\to 0$ and thus $t_n\to\infty$.} As $\{\X^{t_n}\}_{n\in\N}$ is contained in compact set $\bar\cX^2$, there further exists a subsequence $\{t_m'\}_{m'\in\N}\subset \{t_n\}_{n\in\N}$ such that $\X^{t_m'}$ converges to some $\X^\infty\in\bar\cX^2 \subset \cX^\ast$ as $m\to\infty$. Since $\tilde\gamma(t_m')\to 0$ and $\gamma$ is continuous, we have $\gamma(\X^\infty)=0$, which implies by PC of $\v$ that $\X^\infty\in\cX^\ast$ is a stationary point. Since $\X^\ast$ is the only stationary point in $\cX^\ast$, this limit point $\X^\infty$ must be $\X^\ast$. However, as $f(\X^{t_m'})\to \bar f$ and $f$ is also continuous, we have $f(\X^\infty)=\bar f>f(\X^\ast)$  and thus $\X^\infty\ne \X^\ast$, a contradiction.
\end{proof}

\end{document}